\documentclass[aps,pre,twocolumn,notitlepage,citeautoscript,footinbib,superscriptaddress,
eqsecnum]{revtex4-1}  
\usepackage{graphicx}
\usepackage{multirow}
\usepackage{verbatim, bm}
\usepackage{physics}
\usepackage{textcomp}
\usepackage{cancel}
\usepackage{nicefrac}
\usepackage{mathrsfs}  
\usepackage[normalem]{ulem}
\usepackage{mathtools}
\usepackage{amsfonts}

\usepackage{color}
\usepackage{tikz}
\usepackage{tikz-3dplot}
\usepackage{amsmath}
\usepackage{wrapfig}
\usepackage[papersize={8.5in,11in}]{geometry}
\usepackage[colorlinks=true]{hyperref}  
\hypersetup{
    unicode=false,          
    pdftoolbar=true,        
    pdfmenubar=true,        
    pdffitwindow=false,     
    pdfstartview={FitH},    
    pdfsubject={},          
    pdfcreator={},          
    pdfproducer={},         
    pdfkeywords={} {} {},   
    pdfnewwindow=true,      
    colorlinks=true,        
    linkcolor=magenta,      
    citecolor=blue,         
    filecolor=magenta,      
    urlcolor=blue           
} 
\allowdisplaybreaks
\usepackage{mathtools}

\DeclarePairedDelimiter\floor{\lfloor}{\rfloor}
\tdplotsetmaincoords{70}{120} 
\tdplotsetrotatedcoords{0}{0}{0} 
\begin{document}
    \title{Spectral form factors of clean and random quantum Ising chains}
    \author{Nivedita}
    \affiliation{Department of Materials Science, Indian Institute of Technology, Kanpur UP 208016, India}
    \author{Henry Shackleton}
    \affiliation{Department of Physics, Harvard University, Cambridge MA 02138, USA}
    \author{Subir Sachdev}
    \affiliation{Department of Physics, Harvard University, Cambridge MA 02138, USA}
    \date{\today\\\vspace{0.4in}}
    \begin{abstract}
        We compute the spectral form factor of two integrable quantum-critical many body systems in one spatial dimension. The spectral form factor of the quantum Ising chain is periodic in time in the scaling limit described by a conformal field theory; we also compute corrections from lattice effects and deviation from criticality. Criticality in the random Ising chain is described by rare regions associated with a strong randomness fixed point, and these control the long time limit of the spectral form factor.    
    \end{abstract}
    \maketitle
    The spectral form factor (SFF), defined as the Fourier transform of the eigenvalue density-density correlation function, is a useful tool for characterizing spectral statistics in quantum systems. As the behavior of the SFF is well understood for random matrix ensembles~\cite{Mehta2004}, the SFF can be used as an indicator of quantum chaos, where level statistics are predicted to resemble those of random matrices~\cite{Wigner1955, Dyson1962}. This correspondence has been studied in a variety of models, including semiclassical analysis of systems with classically chaotic counterparts~\cite{Berry1985,Sieber2001,Muller2004a,Muller2009}, Floquet systems~\cite{Kos2017,Bertini2018}, holographic systems~\cite{Cotler2016}, systems featuring many-body localization~\cite{Suntajs2019, Halimeh2019}, and for mesoscopic disordered systems~\cite{Efetov1983, Argaman1993}. In the latter case, much work has been done in connecting the spectral statistics of these disordered systems to experimentally-relevant transport properties. All these examples contrast the behavior of integrable models, whose eigenvalue statistics are conjectured to generically be Poissonian~\cite{Berry1977}.

    In this paper, we focus on understanding the universality of the SFF at a quantum critical point (QCP). At a QCP, physical observables such as correlation functions are expected to be described by universal functions of dimensionless parameters. By analogy, one should expect the eigenvalue statistics, and hence the SFF, to be universal as well. One of the simplest examples of a QCP is the one-dimensional quantum Ising model, whose critical point can be mapped to a theory of massless fermions. One can introduce disorder to this model, which drastically alters the behavior at the critical point~\cite{Fisher1995, Motrunich1999}. We study the SFF of both the clean and disordered Ising model at their respective QCPs, obtaining analytic predictions for the scaling behavior in both cases. In the case of the clean Ising model, the correspondence of the critical point with a rational conformal field theory (CFT) leads to a periodic SFF as a function of dimensionless variables. For the disordered critical point, this periodic behavior is replaced by a simple plateau, whose behavior as a function of dimensionless parameters is also universal. These predictions are verified against numerical calculations, and are shown to agree well.
    \section{The Spectral Form Factor and Eigenvalue Statistics}
    For a Hamiltonian $H$ with eigenvalues $E_n$, $n=1,\ldots D$, the normalized eigenvalue density is given by
    \begin{equation}
        \label{eq:eigenvalueDensity}
        \rho\left(E\right)=\frac1D\sum_{n=1}^D\delta\left(E-E_n\right)\,.
    \end{equation}
    The density-density correlation function is given by
    \begin{equation}
        R\left(E_1,E_2\right)=\left\langle\rho\left(E_1\right)\rho\left(E_2\right)\right\rangle
        \label{eq:densityCorrelation}
    \end{equation}
    where brackets are meant to indicate averaging over an ensemble of Hamiltonians.

    The spectral form factor $g(t)$ is given as a Fourier transform of the density-density correlation function
    \begin{equation}
        g(t)=\frac{\int\dd{E_{1,2}}R(E_1,E_2)e^{i((E_1-E_2)t}}{\int\dd{E_{1,2}}R(E_1,E_2)}\,.
        \label{eq:sffNoTemp}
    \end{equation}
    One can generalize this definition by also considering the Fourier transform of the center-of-mass variable, $E_1+E_2$
    \begin{equation}
        g(t,\beta)=\frac{\int\dd{E_{1,2}}R(E_1,E_2)e^{i(E_1-E_2)t-\beta(E_1+E_2)}}{\int\dd{E_{1,2}}\dd{E_2}R(E_1,E_2)e^{-\beta(E_1+E_2)}}\,.
        \label{eq:sffTemp}
    \end{equation}
    Note that this can be written in terms of the partition function, $Z(\beta) = \sum_i e^{-\beta E_i} = \int \dd{E} e^{-\beta E} \rho(E)$
    \begin{equation}
    \begin{aligned}
        g(t, \beta) &= \frac{\langle Z(\beta + i t) Z(\beta - i t) \rangle}{\langle Z(\beta)^2 \rangle} 
        \\
        &= \frac{1}{\langle Z(\beta)^2 \rangle} \Big\langle \sum_{m, n} e^{-\beta (E_m + E_n) + i (E_n - E_m) t} \Big\rangle\,.
        \label{eq:sffPartitionFunction}
        \end{aligned}
    \end{equation}
    Because of this correspondence, we refer to $g(t,\beta)$ as the SFF at inverse temperature $\beta$.

    Often, the density given by Eq.~\ref{eq:eigenvalueDensity} is normalized such that the mean density is constant across the spectrum, referred to as an ``unfolding'' of the eigenvalues. The quantity obtained by analytically continuing the partition function via Eq.~\ref{eq:sffPartitionFunction} lacks this unfolding procedure. For the purposes of this paper, we choose to focus on the quantity defined by Eq.~\ref{eq:sffPartitionFunction}, as it is more amenable to both numerical and analytic studies. This leads to slight discrepancies between the behavior of $g(t,\beta)$ and the one commonly studied in matrix theory, which we will describe below.

    By taking the long-time average of Eq~\ref{eq:sffTemp}, terms with $E_n\neq E_m$ are suppressed, and thus the long-time behavior of $g(t,\beta)$ is predicted to behave for a non-degenerate spectrum as
    \begin{equation}
        \lim_{T \rightarrow \infty} \frac{1}{T} \int_0^T \dd{t} g(t, \beta) = \frac{Z(2\beta)}{Z(\beta)^2} = e^{S(2\beta) - 2 S(\beta)}\,.
        \label{eq:sffAsymptotics}
    \end{equation}
    $S(\beta)$ is the entropy of the system at inverse temperature $\beta$. Not only is the average value of the SFF at long time expected to coincide with Eq.~\ref{eq:sffAsymptotics}, but it approximately plateaus to a constant given by this average value. This is due to the summation of oscillating phases given by  Eq.~\ref{eq:sffPartitionFunction} suppressing $E_n \neq E_m$ terms. The timescale at which this behavior sets in is given by the \text{plateau time}, $t_p$, which scales as the inverse energy spacing.

    There are two typical classes of SFF behavior. The first is indicative of quantum chaos, and arises when eigenvalues repel each other and the eigenvalue spacing is distributed according to the Wigner surmise. This correspondence is well understood for quantum systems with a classically chaotic counterpart~\cite{Berry1985, Sieber2001, Muller2004a, Muller2009}, and is generally taken to be a diagnostic of quantum chaos for systems with no classical limit. In this case, the SFF displays an initial linear ramp with positive slope, ultimately leveling off at a plateau related to the discreteness of the spectrum. A study of the behavior of $g(t,\beta)$ defined in terms of the partition function for a quantum chaotic model was carried out in~\cite{Cotler2016}.

    The second class of systems are integrable. In these systems, eigenvalues are conjectured to generically be uncorrelated and for the eigenvalue spacing to follow a Poisson distribution~\cite{Berry1977}. If this is the case, then
    \begin{equation}
        \begin{aligned}
            g(t, \beta) &= \frac{1}{Z(\beta)^2} \int \dd{E_1} \dd{E_2} \langle \rho(E_1) \rangle \langle \rho(E_2) \rangle 
            \\
            &\times e^{i (E_1 - E_2)t - \beta(E_1 + E_2)} 
            \\
            &= \frac{\langle Z(\beta + i t ) \rangle \langle Z(\beta - i t) \rangle}{Z(\beta)^2}\,.
        \end{aligned}
    \end{equation}
    If the eigenvalue distribution was unfolded ($\langle \rho(E) \rangle = 1$), then $g(t > 0, 0)$ will be constant. Deviations from this constant distribution and finite-temperature effects are reflected by $g(t, \beta)$ instead decaying from its initial value of $1$ to a constant given by Eq.~\ref{eq:sffAsymptotics} over a timescale $t_p$. The lack of eigenvalue repulsion in the level statistics prevents the SFF from dipping below the plateau value~\cite{Cotler2016}.
    \section{SFF of the clean Ising model}
    We now calculate the SFF of the quantum Ising model at criticality, via an explicit calculation of the partition function. The one-dimensional quantum Ising model is defined as
    \begin{equation}
        \mathcal H=-\frac12\sum_{j=0}^{N-1}J_j\sigma_j^z\sigma_{j+1}^z-\frac12\sum_{j=0}^{N-1}g_j\sigma_j^x
        \label{eq:cleanIsingHamiltonian}
    \end{equation}
    where $\sigma^z_j$, $\sigma^x_j$ are Pauli spin operators acting on site $j$, $j= 0,\ldots N-1$. By assuming $\sigma^z_N=\sigma^z_0$, we impose periodic boundary conditions on our system. Note the lack of randomness in this model - this means that the SFF will not involve any sort of averaging procedure.

    Eq.~\ref{eq:cleanIsingHamiltonian} can be mapped via a Jordan-Wigner transformation to a system of free fermions as
    \begin{equation}
        \label{eq:isingFreeFermions}
        \begin{aligned}
            \mathcal H&=\frac12\sum\limits_{j=0}^{N-1}g_j-\sum\limits_{j=0}^{N-1}g_jc_j^\dag c_j
            \\
            &-\frac12\sum_{j=0}^{N-2}J_j\left(c_j^\dag-c_j\right)\left(c_{j+1}^\dag+c_{j+1}\right)\\     &+\frac{J_{N-1}}2e^{i\pi N_f}\left(c_{N-1}^\dag-c_{N-1}\right)\left(c_0^\dag+c_0\right)
        \end{aligned}
    \end{equation}
    where $N_f=\sum_{j=0}^{N-1}c_j^\dag c_j$ is the fermion number. The Hamiltonian commutes with $\exp\left(i\pi N_f\right)$, and thus we can find simultaneous eigenvalues for both operators. Hence, it is enough to solve the system for $\exp\left(i\pi N_f\right)=\pm1$ (its eigenvalues), i.e., solve the above fermion problem for periodic and anti-periodic boundary conditions.
    
    We define $r=\frac{1+\exp\left(i\pi N_f\right)}2$. Then $r=0$ corresponds to periodic and $r=1$ corresponds to anti-periodic boundary conditions. We take the single particle spectrum with periodic and anti-periodic boundary conditions to be $E_k^\text{p}$ and $E_k^\text{ap}$, where $k$ ranges over the respective first Brillouin zones, given by
    \begin{equation*}
        \Lambda_r=\left\{\frac{2\pi}N\left(n-\frac r2\right)\big\vert\; n=0,\dots,N-1\right\}\,.
    \end{equation*}
    Here, we assume a uniform model and take $g_j=g$ and $J_j=J$ for all sites. We set $\left\lvert J\right\rvert=1$ as it simply constitutes an overall rescaling of the energy. As we are
    interested in the thermodynamic limit, we take the system to be
    ferromagnetic ($J>0$), as the antiferromagnetic model is
    frustrated for odd system sizes. The ferromagnetic system in the thermodynamic limit has two distinct gapped phases for different values of $g$, separated by a gapless QCP at $g=1$.

    In the uniform case, for both $r=0,1$, Eq~\ref{eq:isingFreeFermions} can be diagonalised using a Bogoliubov transformation to yield the single particle spectra $E_k^\text{ap}$ and $E_k^\text{p}$ for the antiperiodic and periodic boundary cases. It turns out that both the anti-periodic and periodic systems have the same dispersion relation, only differing in their first Brillouin zones (domains of $k$).
    \begin{equation}
        H_r=\sum_{k\in\Lambda_r}E_k\left(\frac12-d_{k,r}^\dagger d_{k,r}\right)
        \label{eq:diagonalizedIsing}
    \end{equation}
    such that
    \begin{equation}
        E_k=
        \begin{dcases}
            g+1&\text{for }k=0\\
            g-1&\text{for }k=\pi\\
            \sqrt{g^2+2g\cos k+1}&\text{otherwise.}
        \end{dcases}
        \label{eq:isingDispersionRelation}
    \end{equation}
    Note that, although the Ising model is integrable, the linear low-energy dispersion relation at criticality ($E_k\sim k$) leads to a regularly-spaced energy spacing with a large number of degeneracies. As this eigenvalue distribution is neither RMT or Poissonian, we expect an SFF distinct from the two forms discussed earlier.

    The partition function is given by a summation over the many-particle spectrum as
    \begin{equation}
    \begin{aligned}
        Z(\beta)&=\sum_{\{n_k\},\,\text{even}}e^{-\beta\sum_k\left(\nicefrac12-n_k\right) E_k^{\text{ap}}}
        \\
        &+\sum_{\{n_k\},\,\text{odd}}e^{-\beta\sum_k\left(\nicefrac12-n_k\right) E_k^{\text{p}}}
        \label{eq:isingSFF}
        \end{aligned}
    \end{equation}
    where the first (second) summation over many-body states is restricted to those with an even (odd) number of fermions and $k\in\Lambda_1$ ($k\in\Lambda_0$). This constraint can instead be written as
    \begin{equation*}
        \begin{aligned}
            Z(\beta) &= \sum_{\{n_k\}}e^{-\beta\sum_k\left(\nicefrac12-n_k\right)E_k^{\text{ap}}} \delta_{\{n_k\}, \text{even}} 
            \\
            &+\sum_{\{n_k\}} e^{-\beta \sum_k (n_k-1/2) E_k^{\text{p}}}\delta_{\{n_k\}, \text{odd}}\,,
            \\
            \delta_{\{n_k\}, \text{even}}&=\frac12\sum_{\sigma=0,1}e^{-i\pi\sigma\sum_k n_k}\,,
            \\
            \delta_{\{n_k\},\text{odd}}&=\frac12\sum_{\sigma=0,1} e^{-i\pi\sigma\left(\sum_kn_k-1\right)} \,.
        \end{aligned}
    \end{equation*}
    This leads to a simple expression for the partition function in terms of the single particle spectrum:
    \begin{equation}
        \begin{aligned}
            &Z(\beta)=2^{N-1}\times
            \\
            &\Bigg[\prod_{k\in\Lambda_1}\cosh\left(\frac{\beta E^\text{ap}_k}{2}\right)+\left(-1\right)^N\prod_{k\in\Lambda_1}\sinh\left(\frac{\beta E^\text{ap}_k}2\right)\\
            &+\prod_{k\in\Lambda_0}\cosh\left(\frac{\beta E^\text{p}_k}2\right)-\left(-1\right)^N\prod_{k\in\Lambda_0}\sinh\left(\frac{\beta E^\text{p}_k}2\right)\Bigg]\,.
        \label{eq:cleanPartitionFunction}
        \end{aligned}
    \end{equation}
    We now study the behaviour of the partition function after analytic continuation in $\beta$, starting with the universal behavior and then discussing non-universal corrections. The universal behavior of this function at the critical point is obtained by sending $\beta,N\rightarrow\infty$ while holding $\beta/N \equiv z=a+ib$ constant. Since Eq.~\ref{eq:cleanPartitionFunction} is divergent as $N\rightarrow \infty$ due to the extensive ground state energy, we must extract the divergent piece.

    Anticipating universal behavior for small deviations away from the critical point, we take 
    \begin{equation*}
        g=1+\frac\delta N
    \end{equation*}
    and express the partition function in terms of the variables $z,\delta$, and $N$.
    \begin{equation}
    \begin{aligned}
        Z\left(z,\delta,N\right)&=e^{zF\left(\delta,N\right)}\Big[G\left(z,\delta,N\right) +e^{-2z\left\lvert N+\frac\delta2\right\rvert} H\left(z,\delta,N\right)
        \\
        &+\Theta\left(-\left(N+\nicefrac\delta2\right)\right)I\left(z,\delta,N\right)\Big]
        \label{eq:partitionFunctionSimplified}
        \end{aligned}
    \end{equation}
    where
    \begin{widetext}
    \begin{equation*}
        \begin{aligned}
            G(z, \delta, N) &= \cosh \left(\frac{\delta z}2\right) p_0^+(z,\delta, N)^2-\sinh\left(\frac{\delta z}2\right)p_0^-(z, \delta,N)^2+\frac{e^{z\phi(\delta,N)}}2 \left(p_1^+(z,\delta,N)^2+p_1^-(z,\delta, N)^2\right)\\
            F(\delta,N) &= 2 f_0(\delta, N)-\delta_{N, \text{even}}\abs{N+\frac\delta2}\;\;\;\;\;\;\;\;\;\;\;\;\;\;\;\;\;\;\;\phi\left(\delta,N\right)=\left\lvert N+\frac\delta2\right\rvert-2\left(f_0\left(\delta,N\right)-f_1\left(\delta,N\right)\right)\\
            f_r(\delta, N)&=\sum_{n=1}^{\floor{N/2}}\Omega\left(\frac{2\pi}N\left(n-\frac q2\right),\delta,N\right)\;\;\;\;\;\;\;\;\;\;\;\;p_r^s(a,\delta,N)=\prod_{n=1}^{\floor{N/2}}\left(1+s e^{-2z\,\Omega\left(\frac{2\pi}N\left(n-\frac q2\right),\delta,N\right)}\right)\\
            \Omega\left(k,\delta,N\right)&=\sqrt{\frac{\delta^2}4+N(N+\delta) \sin\left(\frac k2\right)^2}
        \end{aligned}
    \end{equation*}
    \end{widetext}
    This is just a rewriting of Equation~\ref{eq:cleanPartitionFunction} in the new variables - no limits have been taken. The exact expressions for $H\left(z,\delta,N\right)$ and $I\left(z,\delta,N\right)$ are given in Eq~\ref{eq:HIexact}. We have not included their full expressions here because, as we will soon argue, they represent non-perturbative corrections to the partition function.
    
    Though this expression for the partition function is complicated, we can make a few observations. Crucially, as shown in Appendix~\ref{ap:CalcCorrections}, all the terms in the brackets of Eq.~\ref{eq:partitionFunctionSimplified} have finite limits as $N\to\infty$. $H\left(z,\delta,N\right)$ is exponentially damped when $\mathrm{Re}\left(z\right)>0$, and $\Theta\left(-\left(N+\nicefrac\delta2\right)\right)=0$ as soon as $N>-\nicefrac\delta2$. Therefore, neither $H\left(z,\delta,N\right)$ nor $I\left(z,\delta,N\right)$ contribute to the partition function at any polynomial order in $\nicefrac1N$.
    
    On the other hand, $F\left(\delta,N\right)$ diverges quadratically in $N$. This is expected, as $\nicefrac{-F\left(\delta,N\right)}N$ is the leading order contribution to the free energy, which itself is extensive and should scale linearly with $N$. The leading contribution to the free energy will come from the ground state, and indeed, $F\left(\delta,N\right)$ is closely related to the ground state energy of the system.
    \begin{equation}
    \begin{aligned}
            -N E_0&=\phi\left(\delta,N\right)+F\left(\delta,N\right)
            \\
            &+\delta_{N,\text{even}}\Theta\left(-\left(2N+\delta\right)\right)\left(2N+\delta\right)
            \\
            &=\phi\left(\delta,N\right)+F\left(\delta,N\right)
        \end{aligned}  
    \end{equation}
    where the second equality holds for $\delta>-2N$. It can be shown from the further expansions of these terms in Appendix~\ref{ap:CalcCorrections} that $E_0= -\nicefrac{2}{\pi}\left(N+\nicefrac \delta2\right) +\mathcal O\left(\delta^2\frac{\log\left(N\right)}N\right)$ 
 
    For the SFF, the overall factor involving $F\left(\delta,N\right)$ will cancel out. Hence, only $G\left(z,\delta,N\right)$ will be relevant for calculating contributions to the SFF. The four terms in $G\left(z,\delta,N\right)$ reflect the four products in Eq~\ref{eq:cleanPartitionFunction}. For $N$ even (odd), the first (second) two terms come from the periodic products, and the other two terms come from the anti-periodic products. The phase $\phi\left(\delta,z\right)$ reflects the difference in the ground state energies of the periodic and anti-periodic fermion system. It is also related to the gap between the ground and first excited state, $\Delta$ \cite{Oshikawa2019UniversalFG}. Precisely, 
    \begin{equation}
        \begin{aligned}
            N\Delta&=(-1)^{\left(1-r\right)\Theta\left(-g\right)}\Big[\phi\left(\delta,N\right)+\frac\delta2
            \\
            &+\left(-1\right)^r\Theta\left(-\left(2N+\delta\right)\right)\left(2N+\delta\right)\Big]
            \\
            &=\phi\left(\delta,N\right)+\frac\delta2
        \end{aligned}
    \end{equation}
    where $r=N+1\mod2$ and the second equality holds when $g>0$ or equivalently, $\delta>-N$. From the above formula and the expression for $\phi\left(\delta,\infty\right)$ we can see that in the thermodynamic limit, the spectrum in Eq.~\ref{eq:diagonalizedIsing} is gapped, with the gap closing at $g = 1$. We can also see that for a finite system size, the critical point has a gap which scales as $\nicefrac 1N$.
    
    The thermodynamic limit of Eq~\ref{eq:partitionFunctionSimplified} and leading-order corrections are worked out carefully in Appendix~\ref{ap:CalcCorrections}. We first consider the thermodynamic limit, where
    \begin{equation*}
        \begin{aligned}
            p_r^s(z,\delta,\infty)&=\prod_{n=1}^\infty\left( 1+s\,e^{-2z\sqrt{\frac{\delta^2}4 + \pi^2 \left(n - \frac r2\right)^2}}\right)\,,\\
            \phi(\delta,\infty) &= \frac{\abs{\delta}}{2} + \frac{1}{\pi} \int_{\abs{\delta}}^\infty \frac{\sqrt{x^2 - \delta^2}}{\sinh(x)} \dd{x}\,.
        \end{aligned}
    \end{equation*}
    Note that this gives an expression for the energy gap in the thermodynamic limit
    \begin{equation}
        N\Delta=\delta\Theta\left(\delta\right)+\frac1\pi\int_{\left\lvert\delta\right\rvert}^\infty\frac{\sqrt{x^2-\delta^2}}{\sinh\left(x\right)}\mathrm dx
    \end{equation}
    in agreement with prior calculations~\cite{Oshikawa2019UniversalFG}. 
    
    At the critical point ($\delta = 0$), this reduces to
      \begin{equation*}
        \begin{aligned}
            p_r^s(z,0,\infty)&=\prod_{n=1}^\infty\left( 1+s\,e^{-2z \pi \left(n - \frac r2\right)}\right)\,,\\
            \phi(0,\infty) &=\frac1\pi \int_0^\infty \frac{x}{\sinh(x)} \dd{x} = \frac{\pi}{4}\,.
        \end{aligned}
    \end{equation*}
    Therefore, the thermodynamic limit of the quantum Ising model at criticality up to $\order{1/N^2}$ corrections is, defining $\tau = iz$,
    \begin{equation}
        Z\left(\tau,0,N\right)=e^{\nicefrac{-2i\tau N^2}\pi}\left(\frac{\mathfrak f\left(\tau\right)^2+\mathfrak f_1\left(\tau\right)^2+\mathfrak f_2\left(\tau\right)^2}2\right)\,,
        \label{eq:isingUniversalPartition}
    \end{equation}
    where $\mathfrak f(\tau)$ are Weber modular functions
    \begin{equation*}
        \begin{aligned}
        \mathfrak f(\tau) &=q^{-\frac1{48}}\prod_{n=1}^\infty\left(1+q^{n-\frac12}\right)=\frac{\eta^2(\tau)}{\eta\left(\frac{\tau}{2}\right) \eta(2\tau)}\,,
        \\
        \mathfrak f_1(\tau) &=q^{-\frac1{48}}\prod_{n=1}^\infty\left(1-q^{n-\frac12}\right)= \frac{\eta\left(\frac{\tau}{2}\right)}{\eta(\tau)}\,,
        \\
        \mathfrak f_2(\tau) &=\sqrt2q^{\frac1{24}}\prod_{n=1}^\infty\left(1+q^n\right)= \frac{\sqrt{2} \eta(2\tau)}{\eta(\tau)}\,,
        \end{aligned}
    \end{equation*}
    with $\eta(\tau)$ the Dedekind eta function,
    \begin{equation*}
        \eta(\tau) = e^{\frac{\pi i \tau}{12}} \prod_{n=1}^\infty \left(1 - e^{2n\pi i \tau}\right)\,.
    \end{equation*}
    From this and the periodic properties of the Weber functions \cite{Zagier97}, one can see that the partition function is invariant under  $\tau\to\tau+24$. However, a translation $\tau\to\tau+8$ only picks up a phase, and hence the full SFF is periodic in $b$ with period $8$. For large $a$, terms in $p_r^s$ with larger frequencies are suppressed, and the SFF is predominantly sinusoidal with period $8$. As $a \rightarrow \infty$, the time dependence in general is suppressed, and $g(a,b) \sim 1$. In the opposite limit, $a \rightarrow 0$, each term in the product of $p_r^s (z,0,\infty)$ contributes equally. These different frequencies contribute destructively except at integer values of $b$, leading to an SFF sharply peaked at integer values of $b$. 
    
    This expression for the partition function at the critical point (after factoring out the diverging piece) is in agreement with the modular invariant partition function of the Ising CFT~\cite{Kudler-Flam2019ConformalDiagnostics}. In fact, the periodic behavior of the SFF at the Ising critical point can be predicted directly from its correspondence with the Ising CFT. In a conformal field theory, the partition function on a torus with lengths $N$ and $\beta$ can be directly written as~\cite{Dyer2017}
    \begin{equation}
        Z(\tau, \bar{\tau}) = \sum_{(h, \bar{h})} N_{h, \bar{h}} q^{h - \frac{c}{24}} \bar{q}^{\bar{h} - \frac{\bar{c}}{24}}
        \label{eq:isingPartitionFunction}
    \end{equation}
    where $q \equiv e^{2\pi i \tau}$, $N_{h, \bar{h}}$ is the degeneracy of states with conformal weight $(h, \bar{h})$, and
    \begin{equation*}
        \tau = \frac{i\beta}{N} \,, \bar{\tau} = -\frac{i\beta}{N}\,.
    \end{equation*}
    This summation includes both primary and descendant fields, and is hence not particularly useful for explicit calculations of the partition function. More practical calculations absorb contributions from irreducible representations of the conformal group into Virasoro characters, with each character corresponding to a primary field. For the Ising CFT with $c = 1/2$ and $h = \bar{h} = 0, 1/2, 1/16$, this recovers the non-divergent piece of Eq.~\ref{eq:isingUniversalPartition}, with the $h=1/16$ character corresponding to the contribution with periodic boundary conditions and the other two characters corresponding to anti-periodic boundary conditions~\cite{DiFrancesco1997}.
    
    In deriving the periodic nature of the SFF, it is simpler to use Eq.~\ref{eq:isingPartitionFunction}, with contributions from both primary and descendent fields. The SFF is obtained by analytically continuing $\beta \rightarrow \beta + i t$, yielding
    \begin{widetext}
    \begin{equation*}
        \begin{aligned}
            &Z(\beta + i t) Z(\beta - i t) = \sum_{h, \bar{h}, g, \bar{g}}N_{h, \bar{h}} N_{g, \bar{g}} \exp\Bigg[ -\frac{2 \pi \beta}{N}\left( h + \bar{h} + g + \bar{g} - \frac{c}{6} \right) + i \frac{2\pi t}{N} (h + \bar{h} - g - \bar{g}) \Bigg]\,.
        \end{aligned}
    \end{equation*}
    \end{widetext}
    From this, one can see that the SFF is periodic in $t/N$ with period $n$ if 
    \begin{equation*}
        \begin{aligned}
            n(h+\bar{h} - g - \bar{g}) \in \mathbb{Z}\, \forall h\,, g\,.
        \end{aligned}
    \end{equation*}
    Since the dimension of descendent fields differ from the dimension of their corresponding primary field by an integer, this condition only needs to be satisfied for the primary fields. This condition can be satisfied for some $n$ for any rational CFT, since all the scaling dimensions of the primary fields are rational numbers. This periodic structure has been previously noted~\cite{Dyer2017, Benjamin2018}, although rational CFTs can still host non-trivial SFFs if one takes $n$ to be large. For the Ising CFT, with primary scaling dimensions $0$, $1/2$, and $1/16$, the SFF is periodic in $t/N$ with period $8$. This is the leading order behavior of the Ising model SFF for large $\beta/N$, although the non-universal behavior is not captured. This argument also applies to other minimal models with corresponding critical points - for example, the three-state Potts model at criticality should have a periodic SFF with a period of $15$.

    We now consider corrections to the partition function, both from finite $N$ and non-zero $\delta$.
    The resulting expression, up to $\order{\frac{1}{N^3}, \frac{\delta}{N^2}, \delta^2}$ corrections, is
    \begin{equation}
    \begin{aligned}
        &Z\left(\tau,\delta,N\right)=e^{i\tau\tilde F\left(\delta,N\right)}\Big[Z_{00}\left(\tau\right)+\delta Z_{10}\left(\tau\right)
        \\
        &+\frac12\left(\frac1{N^2}Z_{02}\left(\tau\right)+2\frac{\delta}NZ_{11}\left(\tau\right)+\delta^2Z_{20}\left(\tau\right)\right)\Big]
        \label{eq:partitionperturbationtheory}
        \end{aligned}
    \end{equation}
    where $\tau=iz$ and
    \begin{equation*}
        \begin{aligned}
            \tilde F\left(\delta,N\right)=&-\nicefrac2\pi\left(N^2+\nicefrac{N\delta}2+\nicefrac{\delta^2\log\left(N\right)}8\right)\\
            Z_{00}\left(\tau\right)=&\frac{\mathfrak f\left(\tau\right)^2+\mathfrak f_1\left(\tau\right)^2+\mathfrak f_2\left(\tau\right)^2}2\\
            Z_{10}\left(\tau\right)=&\frac{i\tau}{2}\eta\left(\tau\right)^2\\
            Z_{02}\left(\tau\right)=&\frac{-i\tau}{4\pi}\overline G_4\left(\tau\right)
            \\
            Z_{11}\left(\tau\right)=&\frac{-i\tau}{4\pi}\overline G_2\left(\tau\right)
            \\
            Z_{20}\left(\tau\right)=&-\frac{i\tau}{4\pi}\overline G_0\left(\tau\right)-\left(\frac{\tau^2}8-\frac{i\tau}{2\pi}\log\left(2\right)\right)\mathfrak f_2\left(\tau\right)^2
        \end{aligned}
    \end{equation*}
    and we define
    \begin{equation*}
        \begin{aligned}
            &\overline G_{2k}\left(\tau\right)=\frac1{2^{2k}}G_{2k}\left(\frac\tau2\right)\left(\mathfrak f\left(\tau\right)^2-\mathfrak f_1\left(\tau\right)^2\right)
            \\
            &+G_{2k}\left(2\tau\right)\left(\mathfrak f\left(\tau\right)^2-\mathfrak f_2\left(\tau\right)^2\right)\\
            &+\frac12G_{2k}\left(\tau\right)\left(\mathfrak f_1\left(\tau\right)^2+\mathfrak f_2\left(\tau\right)^2-\left(1+\frac4{2^{2k}})\right)\mathfrak f\left(\tau\right)^2\right)
        \end{aligned}
    \end{equation*}
    Here, $G_{2k}\left(\tau\right)$ for $k>0$ is the Eisenstein Series of weight $2k$
    \begin{equation*}
        G_{2k}(\tau) = \sum_{(m,n) \in \mathbb{Z}^2\backslash(0,0)} \frac{1}{(m + n \tau)^{2k}}\,.
    \end{equation*}
     We define $G_0\left(\tau\right)= \log\left(\frac{\pi^2}{64}\eta\left(\tau\right)^4\;e^{-\frac{i\pi\tau}3}\right)-2\left(\gamma-1\right)$ to the bring the $Z_{20}$ term to a more familiar expression. This $\log$ is defined such that $G_0\left(\tau\right)$ is continuous and $G_0\left(ia\right)$ is real for real $a$.
    
    We first focus on the finite-size corrections to the partition function, and by analogy the SFF, at criticality. The leading order corrections are proportional to $\tau/N^2$, as the $\bar{G}_4(\tau)$ component is purely oscillatory - hence, deviations from universality will grow in $\tau$ at a rate proportional to $1/N^2$. This prediction holds up to $\tau\sim N^2$, at which point the perturbative expansion begins to break down.

    Away from the critical point ($\delta\neq0$), the leading-order corrections are given by $Z_{10}$. These corrections grow linearly in $\tau$ at a rate proportional to $\delta$. Therefore, we expect
    critical behavior to hold approximately up to timescales $\tau \sim \delta$, at which point deviations become prominent.

\begin{figure*}
        \centering
        \includegraphics[width=0.9\textwidth]{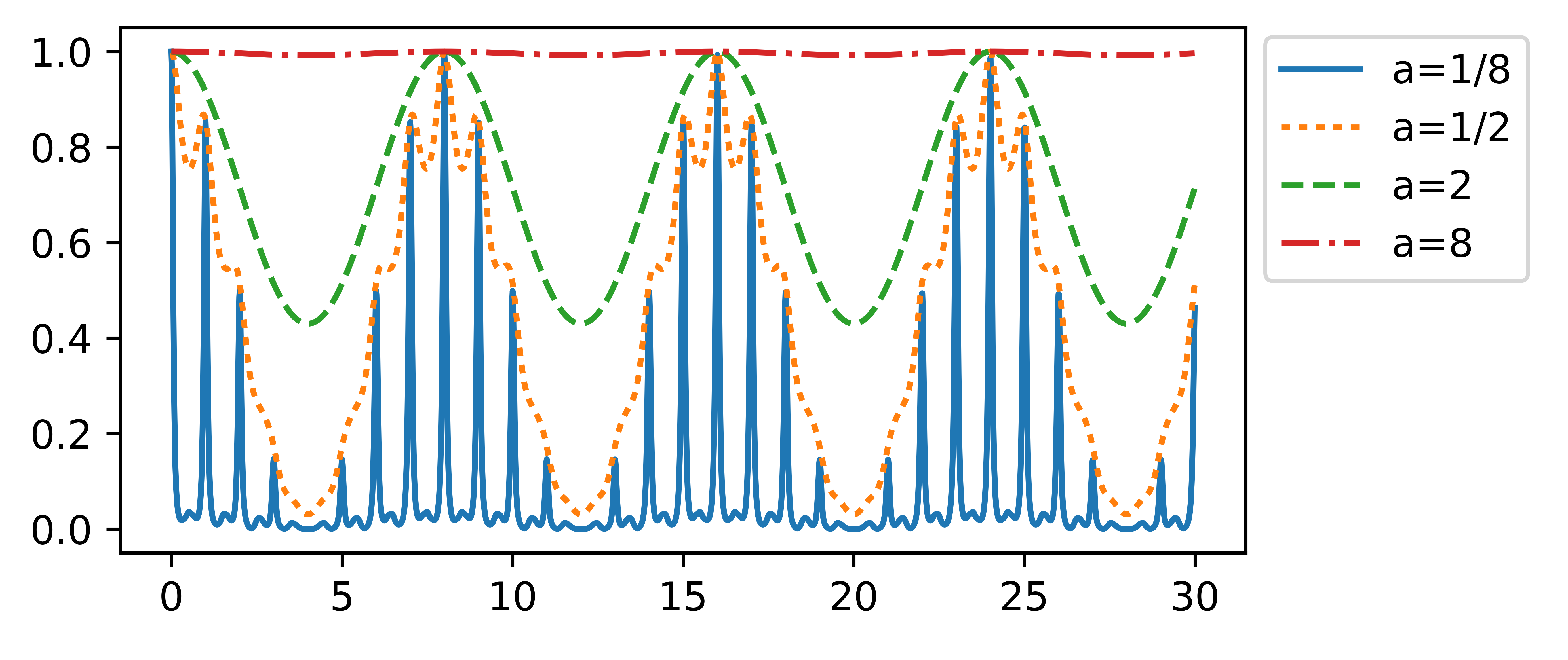}
        \caption{The SFF of the clean Ising model at criticality, plotted as a function of dimensionless parameters $a = \beta/N$ and $b = t/N$, with fixed $N=400$. For all values of $a$, the SFF is periodic in $b$ with a period of $8$. The dominant contribution for large $a$ is a sinusoidal function with period $8$, with higher frequency contributions becoming more prominent with smaller $a$.}
        \label{fig:cleanIsingSFF}
    \end{figure*}

\begin{figure}[ht]
        \centering
        \includegraphics[width=0.45\textwidth]{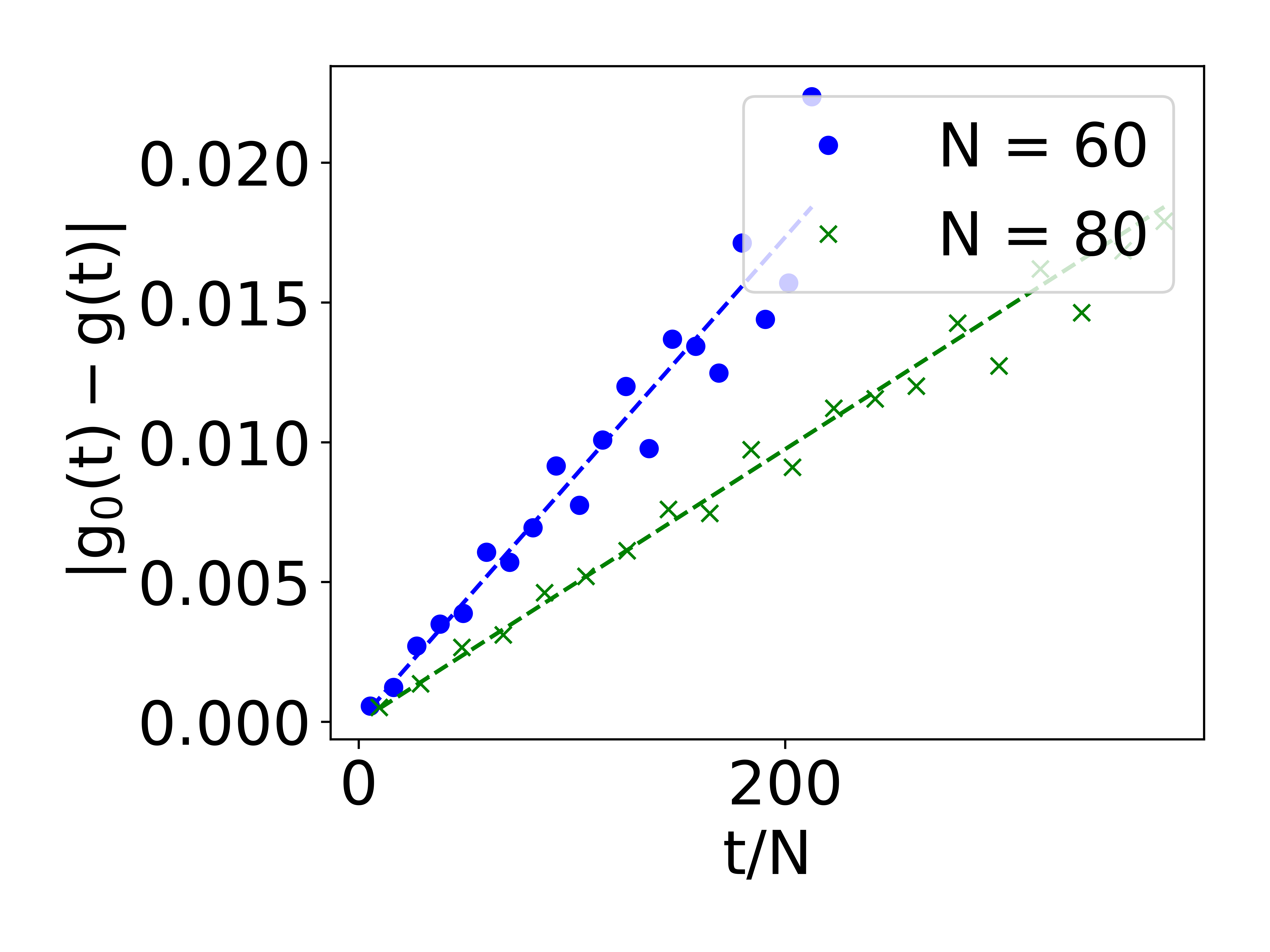}
        \includegraphics[width=0.45\textwidth]{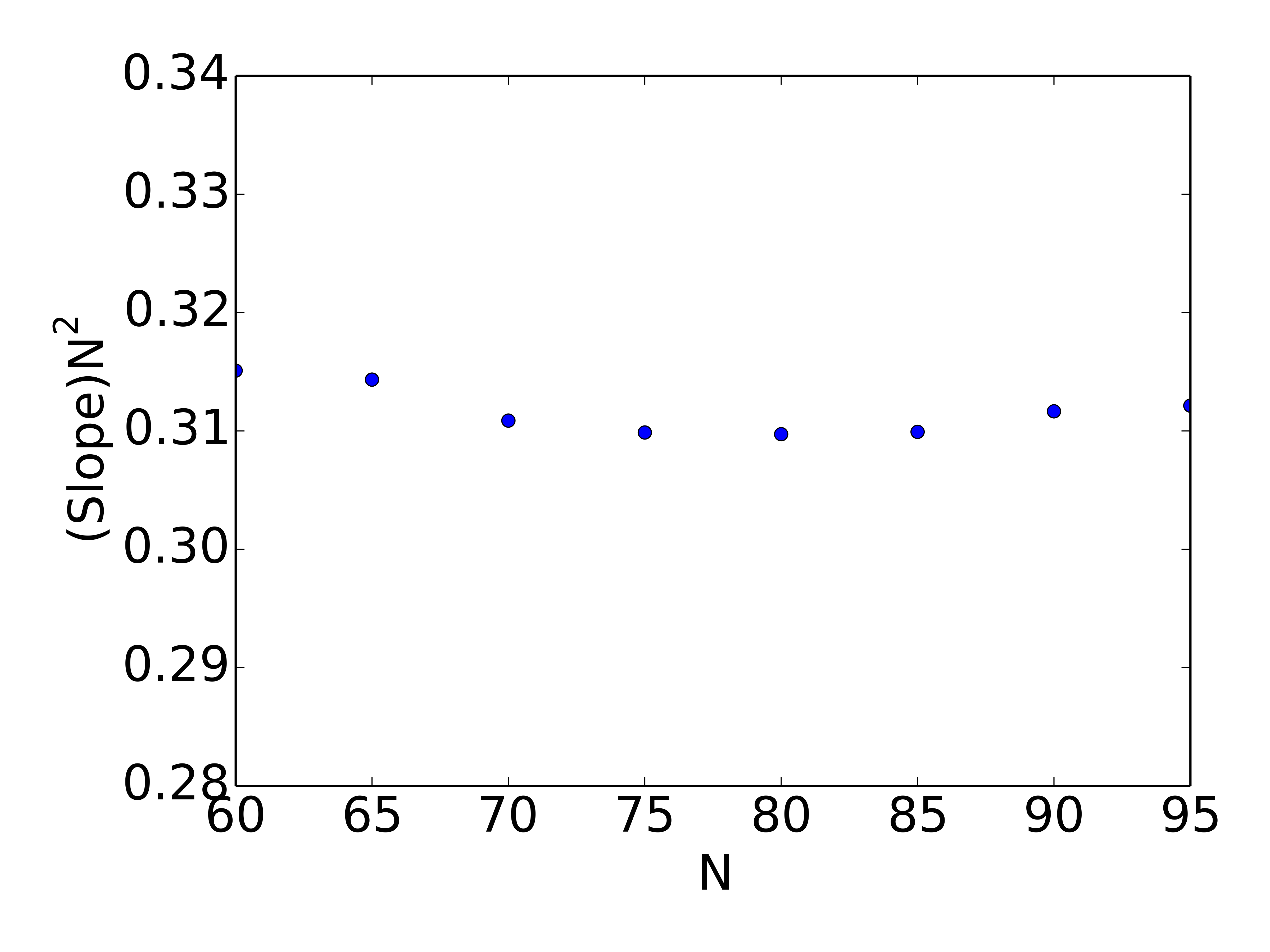}
        \caption{(Top) By numerically calculating the difference between the clean Ising SFF at finite system sizes and the universal behavior predicted analytically, one can see that these deviations grow linearly in $t/N$. The above data was obtained for $a=\beta/N=0.5$, with each point obtained by averaging the deviations across a small window of time, in order to smooth out the data and make the overall linear trend more evident. Analytic predictions are overlaid as dotted lines. (Bottom) The slope of the deviations, multiplied by $N^2$, is constant across a range of system sizes, as predicted by perturbation theory. This constant can also be predicted analytically, as the leading-order correction is of the form $\tau \bar{G}_4(\tau)$, with $\bar{G}_4(\tau)$ oscillatory in time. Averaging over $\abs{\bar{G}_4(\tau)}$ yields an average slope of $\approx 0.312/N^2$ for $a=0.5$, in good agreement with numerical results.}
        \label{fig:nonUniversalCorrections}
    \end{figure}

\begin{figure}[ht]
        \centering
        \includegraphics[width=0.45\textwidth]{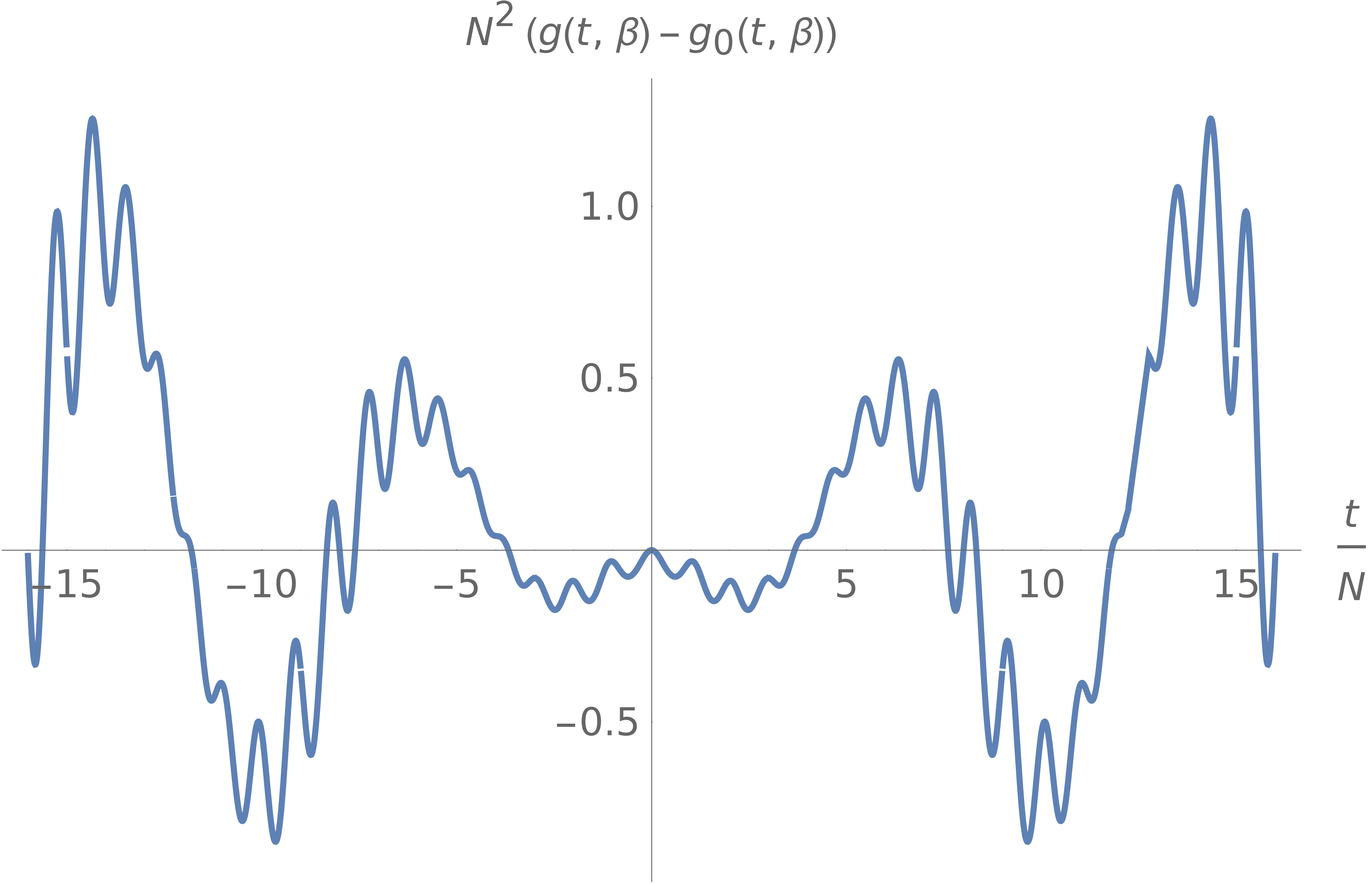}
        \includegraphics[width=0.45\textwidth]{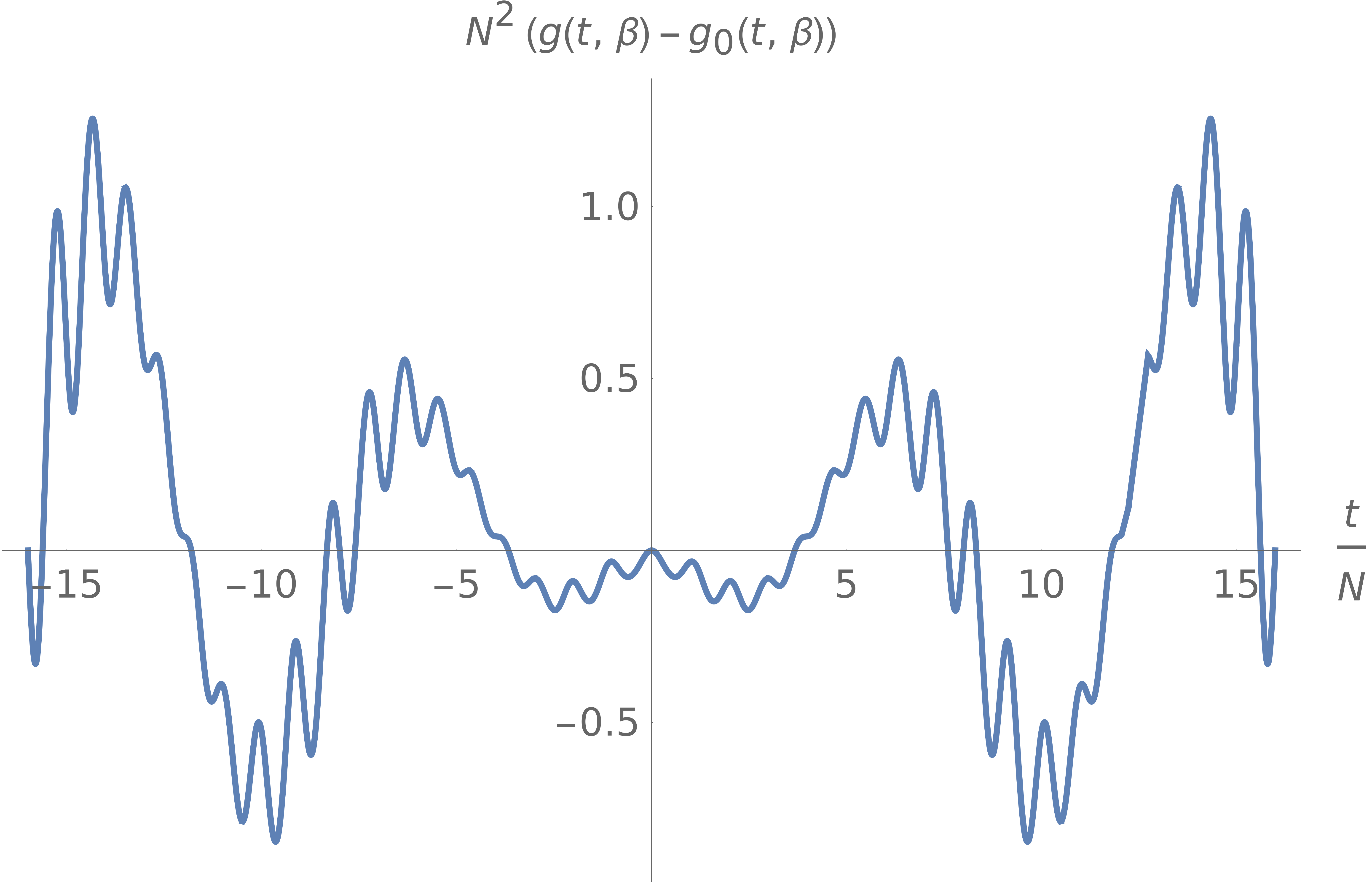}
        \caption{Deviations from the universal behavior of the clean Ising SFF at criticality, with both the numerical results (top) and analytic predictions (bottom). At early times, the two agree well. Parameters used are $N = \beta = 50$.}
        \label{fig:earlyTimeDeviations}
    \end{figure}

    These predictions are verified numerically by calculating the full partition function of the Ising model as given in Eq.~\ref{eq:cleanPartitionFunction}. At criticality, the SFF is indeed periodic in time, with a period and amplitude dependent on the parameters $N$ and $\beta$. As expected from the critical exponent $\nu=1$ for the Ising model (i.e., $N \sim t \sim \beta$), the dominant behavior of the SFF only depends on the dimensionless parameters $a$ and $b$. With this rescaling of time, the SFF is periodic with a period of $8$. The SFF in these rescaled values is shown in Fig.~\ref{fig:cleanIsingSFF}. As predicted by perturbation theory, deviations from this universal behavior will grow linearly in $b$ at a rate proportional to $1/N^2$. This behavior, as well as exact agreement with analytic calculations at early times, is verified in Fig.~\ref{fig:nonUniversalCorrections} and Fig.~\ref{fig:earlyTimeDeviations}.
    \section{SFF of the Disordered Ising Model}
    We now consider the disordered Ising model at criticality, and its corresponding SFF. This model is identical to the clean model given by Eq.~\ref{eq:cleanIsingHamiltonian}, except $J$ and $g$ are replaced with random variables $J_i$ and $g_i$ which vary from site to site. The exact nature of randomness is unimportant; for our purposes, we will take $J_i$ and $g_i$ to be Gaussian distributed with mean $\mu_J$, $\mu_g$ and variances $\sigma_J^2$, $\sigma_g^2$. The SFF of this model has been previously studied for small system sizes~\cite{Lau2018}, although the critical behavior remains unstudied. The introduction of disorder modifies the critical behavior of the Ising model~\cite{Fisher1995}. At the phase transition ($\mu_J = \mu_g = 1$), the system is at an \textit{infinite-randomness fixed point}, where the probability distribution of observables become broadly distributed. 

    The critical behavior of the disordered Ising model is best understood by a disorder renormalization group procedure. This method works by successively eliminating the highest energy terms in the Hamiltonian, leading to an effective low-energy theory. Consider an individual realization of the disordered Ising model, specified by a set of parameters $\{J_i, g_i\}$. We first take the term with the largest energy - either a transverse field $g_i$ or an interaction $J_i$. If the largest term is a transverse field, we set the $i$'th spin to be in its ground state, $\sigma_i^x = 1$, and virtual excitations are treated in perturbation theory which modifies the neighboring interactions. If the largest term is an interaction, the two neighboring spins are combined to form a single, effective spin (cluster) with a modified transverse field. This procedure is repeated, generating a new effective low-energy theory at each step. This leads to a broadening distribution of the effective transverse fields and interactions as the energy scale is lowered. This in turn leads to the RG procedure becoming more accurate, as the largest energy term in each step becomes more likely to be substantially greater than the neighboring terms.

    The precise details of this RG flow were worked out by Fisher~\cite{Fisher1995}. To understand the SFF of the disordered Ising model at criticality, we will cite several relevant results. At an energy scale $\Omega$, the typical density of clusters per unit length is given by
    \begin{equation}
        n \sim \frac{1}{\ln^2 \left( \frac{\Omega_I}{\Omega} \right)}
        \label{eq:disorderedDensity}
    \end{equation}
    where $\Omega_I$ is a UV cutoff given by the largest energy scale of the model. This logarithmic relation between length and energy scales is a general feature of the disordered fixed point, as the average energy gap between the ground and first excited state scales as~\cite{Young1995}
    \begin{equation}
        \Delta E \sim e^{-\sqrt{N}}\,
        \label{eq:disorderedEnergyGap}
    \end{equation}
    as opposed to the $\Delta E \sim 1/N$ behavior in the clean Ising model. Finally, when working at finite system size, it is important to note that the length scale in the RG process at which the effective couplings and transverse fields become broadly distributed is on the order of $1/V_I$, where
    \begin{equation}
        V_I \sim \text{var} (\ln J) + \text{var} (\ln g) \sim \sigma_J^2 + \sigma_g^2\,.
    \end{equation}
    In the thermodynamic limit, any amount of disorder will cause the system to flow to the disordered critical point; for a finite system, one must have $N V_I \gg 1$ in order for behavior to be well-described by the disordered critical point. In our numerical calculations, we verify that the energy gap scaling predicted by Eq.~\ref{eq:disorderedEnergyGap} holds in the parameter regime we consider, which indicates that our disorder is sufficiently strong.

\begin{figure}
        \centering
        \includegraphics[width=0.5\textwidth]{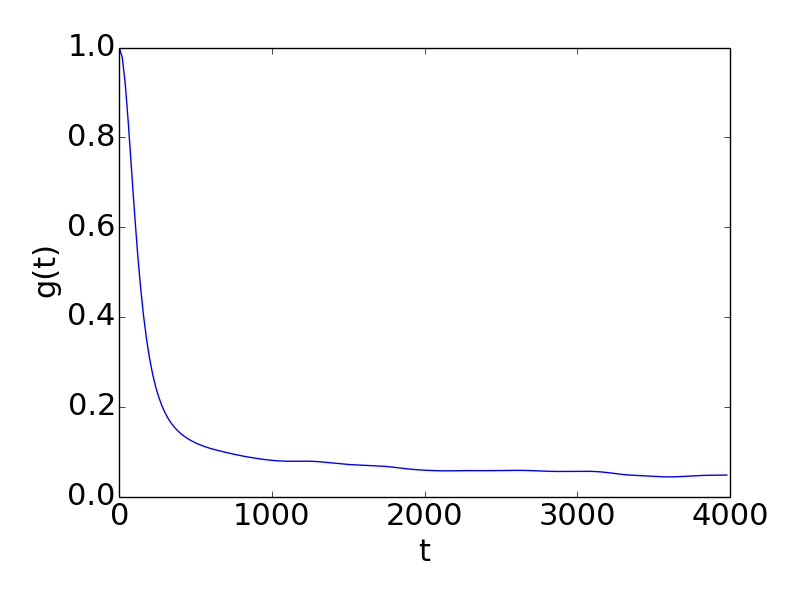}
        \caption{The SFF of the disordered Ising model at criticality, with parameters $N = 400$, $\beta = 200$, $\sigma_J = \sigma_g = 1$. In contrast with the ramp behavior expected in chaotic theories, this SFF quickly decays from its initial value to a constant plateau. This behavior is typical for generic parameter values, so long as $\sigma_{J, g}$ are large enough to bring the finite-size system to the disordered critical point.}
        \label{fig:disorderedIsingSFF}
    \end{figure}

\begin{figure}
        \centering
        \includegraphics[width=0.5\textwidth]{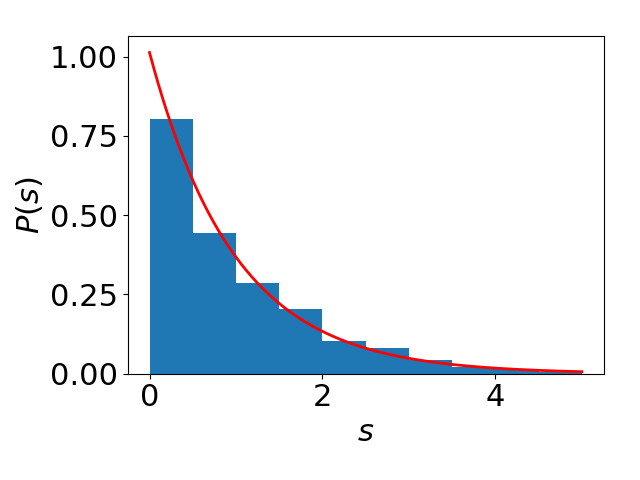}
        \caption{At low energies, the single particle level spacing $s$ of the disordered Ising model at criticality is Poissonian, and shows no indication of eigenvalue repulsion or RMT behavior. Shown is the level spacing and a Poissonian fit for the lowest $1\%$ of single-particle eigenvalues for $N=1000$, $\sigma_J = \sigma_g = 1$.}
        \label{fig:disorderedEigenvalueStatistics}
    \end{figure}

\begin{figure}
        \includegraphics[width=0.5\textwidth]{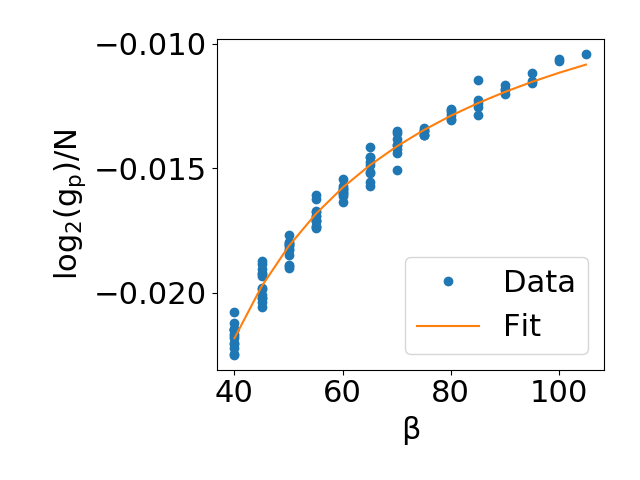}
        \caption{The plateau value of the SFF, computed numerically and fitted to Eq.~\ref{eq:sffPlateauScaling}. The numerical data is obtained for system sizes $40 < N < 120$ and $40 < \beta < N$. $\sigma = 1$ for all data points. The collapse for different values of $N$ is not perfect, indicating additional contributions to the entropy.} 
        \label{fig:sffPlateauFit}
    \end{figure}

    The SFF for the disordered Ising model is evaluated numerically with a method identical to the clean model. For each realization, a Jordan-Wigner transformation is used to convert the system to free fermions, for which a single-particle spectrum is obtained. The partition function is then given by Eq.~\ref{eq:cleanPartitionFunction}. The numerically calculated SFF is shown in Fig.~\ref{fig:disorderedIsingSFF} for a representative system size and temperature. For generic system sizes and temperatures, so long as the disorder is strong enough for Eq.~\ref{eq:disorderedEnergyGap} to hold, the SFF have a simple plateau behavior, with no sign of an RMT ramp. This absence of a ramp suggests Poissonian statistics in the eigenvalue spacing. Indeed, the eigenvalue spacing at low energies, shown in Fig.~\ref{fig:disorderedEigenvalueStatistics}, is Poissonian and no indication of level repulsion is observed for systems sizes up to $N=1000$. This result is not obvious, as the localization length of the eigenstates is known to diverge at zero energy~\cite{Hayn1987, McKenzie1996, Balents1997}. As extended states typically display eigenvalue repulsion, one might expect the SFF to display RMT behavior at low temperatures. This transition from Poissonian to Wigner-Dyson statistics at low energy due to a delocalization transition is present in a number of other disordered systems~\cite{Song1999, Shklovskii1993StatisticsTransition, Jacquod1997}, although all previous examples involve interacting fermions. We attribute the lack of eigenvalue repulsion to the fact that, while the \textit{average} localization length scales as $\xi_a \sim \ln^2 E$, the \textit{typical} localization length, which occurs with probability $1$, scales as $\xi_t \sim \ln E \sim \sqrt{N}$ at low energies~\cite{Balents1997, McKenzie1996}, where we have used Eq.~\ref{eq:disorderedEnergyGap} in the final step. Therefore, while the average localization length may become on the order of the system size, this only arises due to a few anomalous states with large localization length. For a finite system size, one cannot reach a low enough energy such that typical eigenstates extend across the entire system. 
    As our system is at a critical point, we expect the SFF for different parameters to collapse onto a universal function. This can be seen by understanding the behavior of the plateau, which we have previously shown is connected to the entropy via Eq.~\ref{eq:sffAsymptotics}. The entropy of the disordered SFF at finite temperature is calculated by performing the disordered renormalization group procedure up to an energy scale $\Omega = T = \frac{1}{\beta}$. Each free cluster at that energy scale contributes $\ln 2$ to the entropy, leading to an entropy density (using Eq.~\ref{eq:disorderedDensity}) of
    \begin{equation}
        \frac SN\sim\frac1{\ln\left(\beta\Omega_I\right)^2}\,.
        \label{eq:disorderedEntropyDensity}
    \end{equation}
    The plateau value, $g_p$, is then given by
    \begin{equation}
        \begin{aligned}
            \ln g_{p}(\beta) / N &\sim \left[ \ln \left( 2 \beta \Omega_I \right)^{-2} - 2 \ln \left( \beta \Omega_I \right)^{-2} \right]\\
            &\sim \frac{1}{(\ln \beta \Omega_I)^2}\,,
            \label{eq:sffPlateauScaling}
        \end{aligned}
    \end{equation}
    where in the final line, we have assumed $\beta \Omega_I \gg 1$. This scaling behavior agrees well with numerical results, as shown in Fig.~\ref{fig:sffPlateauFit}.
    One should also expect the plateau time $t_p$ to exhibit universal behavior as a function of $\beta$ and $N$. Since the plateau is indicative of the discreteness of the spectrum, we predict $t_p$ to scale inversely with the typical energy spacing. Like the energy gap, this scales as $e^{-\sqrt{N}}$. Additionally, since $t$ and $\beta$ are related to each other via analytic continuation, one would expect $t_b \sim \beta$. Unfortunately, the plateau time does not vary sufficiently as a function of either $N$ or $\beta$ within numerically-accessible ranges to definitively rule out other forms of scaling. In other words, the predicted scaling behavior is indeed consistent with numerical results, but the range of numerically-accessible system sizes is insufficient to see the failure of other forms of scaling, like $t_b \sim N$.
    \begin{figure}
        \centering
        \includegraphics[width=0.45\textwidth]{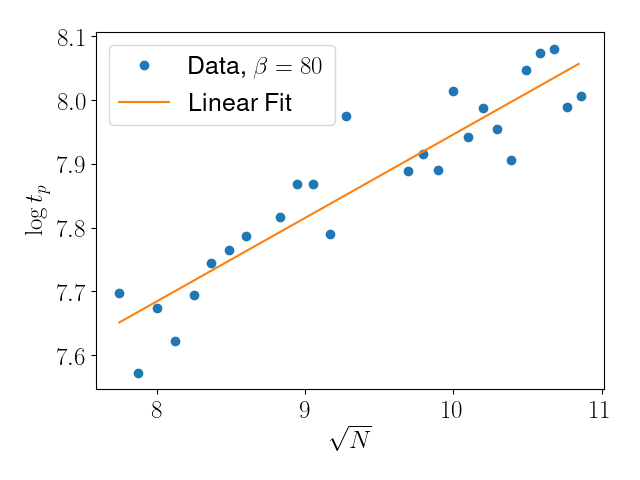}
        \includegraphics[width=0.45\textwidth]{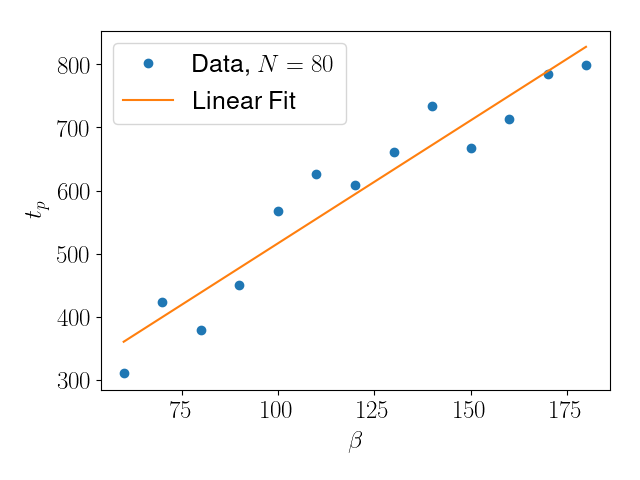}
        \caption{The scaling of the plateau time, $t_p$, of the disordered Ising SFF at criticality, plotted as $\ln t_p$ as a function of $\sqrt{N}$ for fixed $\beta$, and $t_p$ as a function of $\beta$ for fixed $N$. The numerical results are consistent with a $t_p \sim e^{\sqrt{N}} \sim \beta$ scaling; however, the small variance in plateau time is unable to definitely rule out other possible scaling behavior.} 
        \label{fig:plateauTimeScaling}
    \end{figure}
    \section{Conclusions}
    \label{sec:conc}
    The scaling limit of the clean quantum-critical Ising chain is described by a CFT in which all operators have a rational scaling dimension. Consequently the energy levels in a finite-size system are rational numbers, and SFF is a periodic function of time, as shown in Fig.~\ref{fig:cleanIsingSFF}. However, even at the critical point, there are deviations from rationality in a lattice model. These deviations are shown in Fig.~\ref{fig:nonUniversalCorrections} and~\ref{fig:earlyTimeDeviations}: there is excellent agreement between our numerics and the analytic computation of such corrections.

    The critical behavior of the random Ising chain is described by an infinite-randomness fixed point \cite{Fisher1995}. This fixed point specifies the low energy spectrum, and we have shown that it also controls the long-time limit of the SFF. The SFF of this model follows a simple plateau behavior, with no signs of the emergence of an RMT ramp within numerically-accessible system sizes. The infinite-randomness fixed point predicts a universal scaling behavior for the SFF plateau value, which is in good agreement with numerics, as shown in Fig.~\ref{fig:sffPlateauFit}.

    In both these models, the SFF provides valuable insights into the energy level statistics and universality. In the former case, the emergence of a periodic SFF at criticality is indicative of a regular level spacing and an underlying rational CFT description. In the latter case, the simple plateau behavior corresponds to predominantly Poissonian level statistics. In both cases, features of the SFF become universal functions of dimensionless parameters, as is common of observables in critical models.
    \subsection*{Acknowledgements}
    This research was supported by the US Department of Energy Grant DE-SC0019030. Nivedita was supported by a S.N. Bose fellowship from the Department of Science and Technology, Govt. of India, the Indo-U.S. Science and Technology Forum and WINStep Forward.
    \appendix
    \section{Simplification of the Clean Ising Partition Function}
    \label{ap:CalcPart}
    Here, we provide a more explicit calculation of Eq.~\ref{eq:partitionFunctionSimplified} from the original
    form of the partition function.
    Starting from Eq.~\ref{eq:cleanPartitionFunction}, we first extract the $k=0$ and $k=\pi$ modes, as they have a different dispersion. Because $\Omega\left(k,\delta,N\right)$ is symmetric
    about $k = \pi/2$, we can restrict the product over the remaining $k$ to extend from $n=1,\dots,\floor{\nicefrac N2}$. For finite $N$, this leads to slightly different expressions whether $N$ is even or odd. For each case,
    \begin{widetext}
    \begin{equation}
        \begin{aligned}
            Z(z,\delta,N)|_{N\text{odd}}=&e^{2zf_0(\delta,N)} \left(\cosh\left(\frac{\delta z}2\right)p_0^+\left(z, \delta,N\right)^2-\sinh \left(\frac{\delta z}2\right) p_0^-\left(z,\delta,N\right)^2\right)
            +\frac12e^{z\left(\abs{N+\frac{\delta}2}+2f_1(\delta, N)\right)}
            \\
            &\times\Bigg(\left(1+e^{-\abs{2N+\delta}z}\right)p_1^+\left(z,\delta,N\right)^2 +\mathrm{sgn}\left(-\left(N+\nicefrac\delta2\right)\right)\left(1-e^{-\abs{2N+\delta}z}\right)p_1^-(z,\delta, N)^2\Bigg)
            \\
            Z(z,\delta,N)|_{N \text{ even}}=&\frac12e^{2z f_1\left(\delta,N\right)}\left(p_1^+\left(z,\delta,N\right)^2+p_1^-\left(z,\delta,N\right)^2\right)
            +e^{z\left(-\abs{N+\frac\delta2}+2f_0\left(\delta,N\right)\right)}
            \\
            &\times\Bigg(\frac{\cosh\left(\frac{z\delta}2\right)}{1+e^{-z\abs{2N+\delta}}}p_0^+\left(z,\delta,N\right)^2
            -\frac{\mathrm{sgn}\left(-\left(N+\nicefrac\delta2\right)\right)\sinh \left(\frac{z\delta}2\right)}{1-e^{-z\abs{2N+\delta}}}p_0^-\left(z,\delta,N\right)^2\Bigg)
        \end{aligned}
    \end{equation}
    \end{widetext}
    where $f_r\left(\delta,N\right)$ and $p_r^s\left(z,\delta,N\right)$ are as given by Eq~\ref{eq:partitionFunctionSimplified}. The four terms are written in the same order as the products in Eq~\ref{eq:cleanPartitionFunction}. It is straightforward to check that the above partition function can be written as in Eq~\ref{eq:partitionFunctionSimplified} with
    \begin{widetext}
    \begin{equation}
        \begin{aligned}
            H(z,\delta,N) &= 
            \begin{dcases}
                -\left(\dfrac{\cosh\left(\frac{\delta z}2\right)p_0^+(z,\delta,N)^2 }{1+e^{-2z\abs{N+\nicefrac\delta2}}}+\mathrm{sgn}\left(N+\frac\delta2\right)\dfrac{\sinh \left(\frac{\delta z}2\right)p_0^-(z, \delta,N)^2}{{1-e^{-2z\abs{N+\nicefrac\delta2}}}}\right)&\text{for even }N\\
                \dfrac{e^{z\phi(\delta,N)}}2\left(p_1^+(z,\delta, N)^2-p_1^-(z,\delta,N)^2\right)&\text{for odd } N
            \end{dcases}\\
            I(z,\delta,N) &= 
            \begin{dcases}
                2\sinh\left(\frac{\delta z}2\right)p_0^-\left(z,\delta,N\right)^2&\text{for even }N\\
                -e^{z\phi\left(\delta,N\right)}p_1^-\left(z,\delta,N\right)&\text{for odd }N
            \end{dcases}\\
        \end{aligned}
        \label{eq:HIexact}
    \end{equation}
    \end{widetext}
    \section{Corrections to the Partition Function and SFF}
    \label{ap:CalcCorrections}
    At criticality, we calculate the finite-size corrections to the partition function, and hence the SFF.

    We calculate corrections term-wise. First we take $F\left(\delta,N\right)$ and expand in $N$ and $\delta$ to
    second order in $\delta \sim 1/N$, 
    \begin{equation}
        \begin{aligned}
            F\left(\delta,N\right)=&\frac2\pi N^2+\frac{N\delta}\pi-\frac\pi6+\frac{\delta^2\log\left(N\right)}{4\pi}-\frac{\pi^3}{360}\frac1{N^2}
            \\
            &-\frac\pi{12}\frac\delta N+\frac{\log\left(\frac2\pi e^{\gamma-1}\right)}{4\pi}\delta^2 + \order{\frac{1}{N^3}, \delta^3}
        \end{aligned}
    \end{equation}
    This gives corrections to the leading order contribution to the partition function
    \begin{equation}
    \begin{aligned}
        &e^{zF\left(\delta,N\right)}=e^{z\left(\frac{2N^2}\pi-\frac{\pi}6+\frac{N\delta}{\pi}+\frac{\delta^2\log\left(N\right)}{4\pi}\right)}\Bigg[1+
        \\
        &z\left(-\frac{\pi^3}{360}\frac1{N^2}-\frac\pi{12}\frac\delta N+\frac {\log\left(\frac2\pi e^{\gamma-1}\right)}{4\pi}\delta^2\right)\Bigg]
        \end{aligned}
    \end{equation}
    Next, we expand the phase $\phi\left(\delta,N\right)$ 
    \begin{equation}
        \begin{aligned}
            &\phi\left(\delta,N\right)=\left\lvert N+\frac\delta2\right\rvert-2\left(f_0\left(\delta,N\right)-f_1\left(\delta,N\right)\right)\\
            =&\sum_{n=1}^{2N-1}\left(-1\right)^{n-1}\Omega\left(\frac{\pi n}N,\delta,N\right)\\
            =&N\tan\left(\frac\pi{4N}\right)+\frac\delta2\tan\left(\frac\pi{4N}\right)+\frac{\log\left(2\right)}{2\pi}\delta^2
        \end{aligned}
        \label{eq:expansionofphi}
    \end{equation}
    Taylor expanding again gives us the leading order corrections to $e^{z\phi}$,
    \begin{equation}
        e^{z\phi\left(\delta,N\right)}=e^{\frac{\pi z}4}\left[]1+z\left(\frac1{N^2}\frac{\pi^3}{192}+\frac\delta N\frac\pi8+\delta^2\frac{\log\left(2\right)}{2\pi}\right)\right]
    \end{equation}
    For $G\left(z,\delta,N\right)$ we need to evaluate the finite-$N$
    and non-zero $\delta$ corrections of the constituent functions
    \begin{equation*}
        p_r^s(a,\delta,N)=\prod_{n=1}^{\floor{N/2}}\left(1+s e^{-2z\,\Omega\left(\frac{2\pi}N\left(n-\frac q2\right),\delta,N\right)}\right)
    \end{equation*}
    An important observation for analyzing these functions is that
    the $N\rightarrow \infty$ in the thermodynamic limit can be separated into two separate limits - one that takes $N$ dependence
    of the product to infinity, and another than takes the $N$ dependence in the exponential to infinity:
    \begin{equation}
        \lim\limits_{N\to\infty}p_r^s\left(z,\delta,N\right)=\lim\limits_{n_0, N\to\infty}\prod_{n=1}^{n_0}\left(1+s\, e^{-2z \Omega\left(\frac{\pi}N\left(n-\frac r2\right),\delta,N\right)}\right)
    \end{equation}
    The $n_0$ limit converges exponentially fast. Therefore, we will find all the perturbative finite-$N$ corrections from the $N$ limit, and the corrections to the $n_0 \rightarrow \infty$ limit will have no effect in perturbation theory. This justifies the Taylor expansions in the the following steps. 
    
    We first expand $e^{-2z\Omega}$ to obtain the leading order corrections,
    \begin{equation}
    \begin{aligned}
        &s\,e^{-2z\Omega\left(\delta,N\right)}=s\,e^{-2\pi z\left(n-\frac r2\right)}\Bigg[1+
        z\Bigg(\frac1{N^2}\frac{\pi^3\left(n-\frac r2\right)^3}3
        \\
        &-\frac\delta N\pi\left(n-\frac r2\right)-\delta^2\frac1{4\pi\left(n-\frac r2\right)}\Bigg)\Bigg]
        \\
        &\equiv A_n \left[1 + B_n\right]
        \label{eq:energyExpansion}
        \end{aligned}
    \end{equation}
    where $A_n=s\,e^{-2\pi z\left(n-\nicefrac r2\right)}$. Then
    we can expand $p_r^s(z, \delta, N)$ to leading order
    \begin{equation}
        \begin{split}
            &p_r^s(z,\delta,N)=\prod_{n=1}^{\left\lfloor\nicefrac N2\right\rfloor}\left(1+A_n\right)+\sum_{m=1}^{\left\lfloor\nicefrac N2\right\rfloor}\left(B_m\prod_{\substack{n=1
            \\n\neq m}}^{\left\lfloor\nicefrac N2\right\rfloor}\left(1+A_n\right)\right)\\
            =&\prod_{n=1}^{\left\lfloor\nicefrac N2\right\rfloor}\left(1+A_n\right)\left(1+\sum_{m=1}^{\left\lfloor\nicefrac N2\right\rfloor}\frac{B_m}{1+A_m}\right)\\
            =&\prod_{n=1}^\infty\left(1+A_n\right)\left(1+\sum_{m=1}^\infty\frac{B_m}{1+A_m}\right)\\
            =&p_r^s\left(z,0,\infty\right)\Big[1+z\Big(\frac1{N^2}\frac{\pi^3}3R_{r,3}^s\left(z\right)
            \\
            &-\frac\delta N\pi R_{r,1}^s\left(z\right)-\delta^2\frac1{4\pi}R_{r,-1}^s\left(z\right)\Big)\Big]
        \end{split}
    \end{equation}
    where in the second last step, we replaced all upper bounds by $\infty$ as they would give exponentially small corrections, and
    \begin{equation}
        R^s_{r,t}\left(z\right)=\sum_{n=1}^\infty\frac{\left(n-\frac{r}{2}\right)^t}{1+s\,e^{2\pi z\left(n-\frac{r}{2}\right)}},.
    \end{equation}
    One can note that
    \begin{equation}
        \begin{split}
            R_{1,t}^s\left(z\right)+R_{0,t}^s\left(z\right)     &=\frac1{2^t}R_{0,t}^s\left(\frac z2\right)\\
            R_{r,t}^+\left(z\right)+R_{r,t}^-\left(z\right)&=2R_{r,t}^-\left(2z\right)
        \end{split}
    \end{equation}
    Next, we re-express $p_j^s\left(z,0,\infty\right)$ and $R_j^s\left(z\right)$ in terms of modular forms and related quantities. We have
    \begin{equation}
        \begin{aligned}
            p_0^-\left(z,0,\infty\right)&=q^{-\frac1{24}}\eta\left(\tau\right)\\
            p_0^+\left(z,0,\infty\right)&=\frac{q^{-\frac1{24}}}{\sqrt2}\mathfrak f_2\left(\tau\right)\\
            p_1^-\left(z,0,\infty\right)&=q^{\frac1{48}}\mathfrak f_1\left(\tau\right)\\
            p_1^+\left(z,0,\infty\right)&=q^{\frac1{48}}\mathfrak f\left(\tau\right)
            \\
            R_{0,3}^-\left(z\right)&=\frac{1}{240}\left(1-\frac{45}{\pi^4}G_4\left(\tau\right)\right)\\
            R_{0,1}^-\left(z\right)&=-\frac{1}{24}\left(1-\frac3{\pi^2}G_2\left(\tau\right)\right)\\
            R_{0,-1}^-\left(z\right)&=\log\left(q^{-\frac1{24}}\;\eta\left(\tau\right)\right)\\
        \end{aligned}
    \end{equation}
    where $\tau=iz$ and $q=e^{2\pi i\tau}$. In the last expression for $R_{0,-1}^-\left(z\right)$ the $\log$ is chosen such that $R_{0,-1}^-\left(z\right)$ continuous and is real for real $z$. Plugging in these expressions to the definition of $G\left(z,\delta,N\right)$, multiplying with the term $e^{z\,F\left(\delta,N\right)}$ and finally expanding to the relevant orders, we obtain the perturbation theory expression for the partition function mentioned in Eq~\ref{eq:partitionperturbationtheory}.  
     
    With this, the leading order corrections to the SFF can be calculated. We define, 
    \begin{equation}
    \begin{aligned}
        \tilde g_{\infty}\left(a\right)&=\frac{Z_{00}\left(ia-b\right)}{Z_{00}\left(ia\right)}
        \\
        \tilde g_{ij}\left(a,b\right)&= \frac{Z_{ij}\left(ia-b\right)}{Z_{00}\left(ia-b\right)}- \frac{Z_{ij}\left(ia\right)}{Z_{00}\left(ia\right)}
        \\
        \tilde h_{ij}\left(a\right)&=\frac{Z_{ij}\left(ia\right)}{Z_{00}\left(ia\right)}
        \end{aligned}
    \end{equation}
    In terms of these parameters, the corrections to the SFF are given by
    \begin{widetext}
    \begin{equation}
        \begin{aligned}
            g\left(a,b,\delta,N\right)=\left\lvert\tilde g_\infty\left(a,b\right)\right\rvert^2\Bigg(1&+2\mathrm{Re}\left(\tilde g_{10}\left(a,b\right)\right)\delta+2\mathrm{Re}\left(\tilde g_{11}\left(a,b\right)\right)\frac\delta N            +\mathrm{Re}\left(\tilde g_{02}\left(a,b\right)\right)\frac1{N^2}\\
            &+\delta^2\left(\mathrm{Re}\left(\tilde g_{20}\left(a,b\right)\right)+\left\lvert\tilde g_{10}\left(a,b\right)\right\rvert^2-2h_{10}\left(a\right)\mathrm{Re}\left(\tilde g_{10}\left(a,b\right)\right)\right)\Bigg)
        \end{aligned}
    \end{equation}
    \end{widetext}
    Agreement between this complicated expression and the numerically-evaluated SFF has been verified, and confirmed to hold for early times.
    \bibliography{references}

\begin{thebibliography}{31}%
\makeatletter
\providecommand \@ifxundefined [1]{%
 \@ifx{#1\undefined}
}%
\providecommand \@ifnum [1]{%
 \ifnum #1\expandafter \@firstoftwo
 \else \expandafter \@secondoftwo
 \fi
}%
\providecommand \@ifx [1]{%
 \ifx #1\expandafter \@firstoftwo
 \else \expandafter \@secondoftwo
 \fi
}%
\providecommand \natexlab [1]{#1}%
\providecommand \enquote  [1]{``#1''}%
\providecommand \bibnamefont  [1]{#1}%
\providecommand \bibfnamefont [1]{#1}%
\providecommand \citenamefont [1]{#1}%
\providecommand \href@noop [0]{\@secondoftwo}%
\providecommand \href [0]{\begingroup \@sanitize@url \@href}%
\providecommand \@href[1]{\@@startlink{#1}\@@href}%
\providecommand \@@href[1]{\endgroup#1\@@endlink}%
\providecommand \@sanitize@url [0]{\catcode `\\12\catcode `\$12\catcode
  `\&12\catcode `\#12\catcode `\^12\catcode `\_12\catcode `\%12\relax}%
\providecommand \@@startlink[1]{}%
\providecommand \@@endlink[0]{}%
\providecommand \url  [0]{\begingroup\@sanitize@url \@url }%
\providecommand \@url [1]{\endgroup\@href {#1}{\urlprefix }}%
\providecommand \urlprefix  [0]{URL }%
\providecommand \Eprint [0]{\href }%
\providecommand \doibase [0]{http://dx.doi.org/}%
\providecommand \selectlanguage [0]{\@gobble}%
\providecommand \bibinfo  [0]{\@secondoftwo}%
\providecommand \bibfield  [0]{\@secondoftwo}%
\providecommand \translation [1]{[#1]}%
\providecommand \BibitemOpen [0]{}%
\providecommand \bibitemStop [0]{}%
\providecommand \bibitemNoStop [0]{.\EOS\space}%
\providecommand \EOS [0]{\spacefactor3000\relax}%
\providecommand \BibitemShut  [1]{\csname bibitem#1\endcsname}%
\let\auto@bib@innerbib\@empty
\bibitem [{\citenamefont {Mehta}(1991)}]{Mehta2004}%
  \BibitemOpen
  \bibfield  {author} {\bibinfo {author} {\bibfnamefont {M.~L.}\ \bibnamefont
  {Mehta}},\ }\href {\doibase 10.1016/c2009-0-22297-5} {\emph {\bibinfo {title}
  {Random Matrices}}}\ (\bibinfo  {publisher} {Academic Press},\ \bibinfo
  {year} {1991})\ p.\ \bibinfo {pages} {688}\BibitemShut {NoStop}%
\bibitem [{\citenamefont {Wigner}(1955)}]{Wigner1955}%
  \BibitemOpen
  \bibfield  {author} {\bibinfo {author} {\bibfnamefont {E.~P.}\ \bibnamefont
  {Wigner}},\ }\bibfield  {title} {\enquote {\bibinfo {title} {{Characteristic
  Vectors of Bordered Matrices With Infinite Dimensions}},}\ }\href {\doibase
  10.2307/1970079} {\bibfield  {journal} {\bibinfo  {journal} {The Annals of
  Mathematics}\ }\textbf {\bibinfo {volume} {62}},\ \bibinfo {pages} {548}
  (\bibinfo {year} {1955})}\BibitemShut {NoStop}%
\bibitem [{\citenamefont {Dyson}(1962)}]{Dyson1962}%
  \BibitemOpen
  \bibfield  {author} {\bibinfo {author} {\bibfnamefont {F.~J.}\ \bibnamefont
  {Dyson}},\ }\bibfield  {title} {\enquote {\bibinfo {title} {{Statistical
  theory of the energy levels of complex systems. I}},}\ }\href {\doibase
  10.1063/1.1703773} {\bibfield  {journal} {\bibinfo  {journal} {Journal of
  Mathematical Physics}\ }\textbf {\bibinfo {volume} {3}},\ \bibinfo {pages}
  {140} (\bibinfo {year} {1962})}\BibitemShut {NoStop}%
\bibitem [{\citenamefont {Berry}(1985)}]{Berry1985}%
  \BibitemOpen
  \bibfield  {author} {\bibinfo {author} {\bibfnamefont {M.~V.}\ \bibnamefont
  {Berry}},\ }\bibfield  {title} {\enquote {\bibinfo {title} {{Semiclassical
  Theory of Spectral Rigidity}},}\ }\href {\doibase 10.1098/rspa.1985.0078}
  {\bibfield  {journal} {\bibinfo  {journal} {Proc. Roy. Soc. A}\ }\textbf
  {\bibinfo {volume} {400}},\ \bibinfo {pages} {229} (\bibinfo {year}
  {1985})}\BibitemShut {NoStop}%
\bibitem [{\citenamefont {Sieber}\ and\ \citenamefont
  {Richter}(2001)}]{Sieber2001}%
  \BibitemOpen
  \bibfield  {author} {\bibinfo {author} {\bibfnamefont {M.}~\bibnamefont
  {Sieber}}\ and\ \bibinfo {author} {\bibfnamefont {K.}~\bibnamefont
  {Richter}},\ }\bibfield  {title} {\enquote {\bibinfo {title} {{Correlations
  between periodic orbits and their r{\^{o}}le in spectral statistics}},}\
  }\href {\doibase 10.1238/physica.topical.090a00128} {\bibfield  {journal}
  {\bibinfo  {journal} {Physica Scripta T}\ }\textbf {\bibinfo {volume} {90}},\
  \bibinfo {pages} {128} (\bibinfo {year} {2001})}\BibitemShut {NoStop}%
\bibitem [{\citenamefont {{M{\"u}ller}}\ \emph {et~al.}(2004)\citenamefont
  {{M{\"u}ller}}, \citenamefont {{Heusler}}, \citenamefont {{Braun}},
  \citenamefont {{Haake}},\ and\ \citenamefont {{Altland}}}]{Muller2004a}%
  \BibitemOpen
  \bibfield  {author} {\bibinfo {author} {\bibfnamefont {S.}~\bibnamefont
  {{M{\"u}ller}}}, \bibinfo {author} {\bibfnamefont {S.}~\bibnamefont
  {{Heusler}}}, \bibinfo {author} {\bibfnamefont {P.}~\bibnamefont {{Braun}}},
  \bibinfo {author} {\bibfnamefont {F.}~\bibnamefont {{Haake}}}, \ and\
  \bibinfo {author} {\bibfnamefont {A.}~\bibnamefont {{Altland}}},\ }\bibfield
  {title} {\enquote {\bibinfo {title} {{Semiclassical Foundation of
  Universality in Quantum Chaos}},}\ }\href {\doibase
  10.1103/PhysRevLett.93.014103} {\bibfield  {journal} {\bibinfo  {journal}
  {Phys. Rev. Lett.}\ }\textbf {\bibinfo {volume} {93}},\ \bibinfo {eid}
  {014103} (\bibinfo {year} {2004})},\ \Eprint
  {http://arxiv.org/abs/nlin/0401021} {arXiv:nlin/0401021} \BibitemShut
  {NoStop}%
\bibitem [{\citenamefont {{M{\"u}ller}}\ \emph {et~al.}(2009)\citenamefont
  {{M{\"u}ller}}, \citenamefont {{Heusler}}, \citenamefont {{Altland}},
  \citenamefont {{Braun}},\ and\ \citenamefont {{Haake}}}]{Muller2009}%
  \BibitemOpen
  \bibfield  {author} {\bibinfo {author} {\bibfnamefont {S.}~\bibnamefont
  {{M{\"u}ller}}}, \bibinfo {author} {\bibfnamefont {S.}~\bibnamefont
  {{Heusler}}}, \bibinfo {author} {\bibfnamefont {A.}~\bibnamefont
  {{Altland}}}, \bibinfo {author} {\bibfnamefont {P.}~\bibnamefont {{Braun}}},
  \ and\ \bibinfo {author} {\bibfnamefont {F.}~\bibnamefont {{Haake}}},\
  }\bibfield  {title} {\enquote {\bibinfo {title} {{Periodic-orbit theory of
  universal level correlations in quantum chaos}},}\ }\href {\doibase
  10.1088/1367-2630/11/10/103025} {\bibfield  {journal} {\bibinfo  {journal}
  {New Journal of Physics}\ }\textbf {\bibinfo {volume} {11}},\ \bibinfo {eid}
  {103025} (\bibinfo {year} {2009})},\ \Eprint {http://arxiv.org/abs/0906.1960}
  {arXiv:0906.1960 [nlin.CD]} \BibitemShut {NoStop}%
\bibitem [{\citenamefont {Kos}\ \emph {et~al.}(2018)\citenamefont {Kos},
  \citenamefont {Ljubotina},\ and\ \citenamefont {Prosen}}]{Kos2017}%
  \BibitemOpen
  \bibfield  {author} {\bibinfo {author} {\bibfnamefont {P.}~\bibnamefont
  {Kos}}, \bibinfo {author} {\bibfnamefont {M.}~\bibnamefont {Ljubotina}}, \
  and\ \bibinfo {author} {\bibfnamefont {T.}~\bibnamefont {Prosen}},\
  }\bibfield  {title} {\enquote {\bibinfo {title} {{Many-body quantum chaos:
  Analytic connection to random matrix theory}},}\ }\href {\doibase
  10.1103/PhysRevX.8.021062} {\bibfield  {journal} {\bibinfo  {journal} {Phys.
  Rev. X}\ }\textbf {\bibinfo {volume} {8}},\ \bibinfo {pages} {021062}
  (\bibinfo {year} {2018})},\ \Eprint {http://arxiv.org/abs/1712.02665}
  {arXiv:1712.02665 [nlin.CD]} \BibitemShut {NoStop}%
\bibitem [{\citenamefont {Bertini}\ \emph {et~al.}(2018)\citenamefont
  {Bertini}, \citenamefont {Kos},\ and\ \citenamefont {Prosen}}]{Bertini2018}%
  \BibitemOpen
  \bibfield  {author} {\bibinfo {author} {\bibfnamefont {B.}~\bibnamefont
  {Bertini}}, \bibinfo {author} {\bibfnamefont {P.}~\bibnamefont {Kos}}, \ and\
  \bibinfo {author} {\bibfnamefont {T.}~\bibnamefont {Prosen}},\ }\bibfield
  {title} {\enquote {\bibinfo {title} {{Exact Spectral Form Factor in a Minimal
  Model of Many-Body Quantum Chaos}},}\ }\href {\doibase
  10.1103/PhysRevLett.121.264101} {\bibfield  {journal} {\bibinfo  {journal}
  {Phys. Rev. Lett.}\ }\textbf {\bibinfo {volume} {121}},\ \bibinfo {pages}
  {264101} (\bibinfo {year} {2018})},\ \Eprint
  {http://arxiv.org/abs/1805.00931} {arXiv:1805.00931 [nlin.CD]} \BibitemShut
  {NoStop}%
\bibitem [{\citenamefont {Cotler}\ \emph {et~al.}(2017)\citenamefont {Cotler},
  \citenamefont {Gur-Ari}, \citenamefont {Hanada}, \citenamefont {Polchinski},
  \citenamefont {Saad}, \citenamefont {Shenker}, \citenamefont {Stanford},
  \citenamefont {Streicher},\ and\ \citenamefont {Tezuka}}]{Cotler2016}%
  \BibitemOpen
  \bibfield  {author} {\bibinfo {author} {\bibfnamefont {J.~S.}\ \bibnamefont
  {Cotler}}, \bibinfo {author} {\bibfnamefont {G.}~\bibnamefont {Gur-Ari}},
  \bibinfo {author} {\bibfnamefont {M.}~\bibnamefont {Hanada}}, \bibinfo
  {author} {\bibfnamefont {J.}~\bibnamefont {Polchinski}}, \bibinfo {author}
  {\bibfnamefont {P.}~\bibnamefont {Saad}}, \bibinfo {author} {\bibfnamefont
  {S.~H.}\ \bibnamefont {Shenker}}, \bibinfo {author} {\bibfnamefont
  {D.}~\bibnamefont {Stanford}}, \bibinfo {author} {\bibfnamefont
  {A.}~\bibnamefont {Streicher}}, \ and\ \bibinfo {author} {\bibfnamefont
  {M.}~\bibnamefont {Tezuka}},\ }\bibfield  {title} {\enquote {\bibinfo {title}
  {{Black Holes and Random Matrices}},}\ }\href {\doibase
  10.1007/JHEP09(2018)002, 10.1007/JHEP05(2017)118} {\bibfield  {journal}
  {\bibinfo  {journal} {JHEP}\ }\textbf {\bibinfo {volume} {05}},\ \bibinfo
  {pages} {118} (\bibinfo {year} {2017})},\ \bibinfo {note} {[Erratum:
  JHEP09,002(2018)]},\ \Eprint {http://arxiv.org/abs/1611.04650}
  {arXiv:1611.04650 [hep-th]} \BibitemShut {NoStop}%
\bibitem [{\citenamefont {{{\v{S}}untajs}}\ \emph {et~al.}(2019)\citenamefont
  {{{\v{S}}untajs}}, \citenamefont {{Bon{\v{c}}a}}, \citenamefont {{Prosen}},\
  and\ \citenamefont {{Vidmar}}}]{Suntajs2019}%
  \BibitemOpen
  \bibfield  {author} {\bibinfo {author} {\bibfnamefont {J.}~\bibnamefont
  {{{\v{S}}untajs}}}, \bibinfo {author} {\bibfnamefont {J.}~\bibnamefont
  {{Bon{\v{c}}a}}}, \bibinfo {author} {\bibfnamefont {T.}~\bibnamefont
  {{Prosen}}}, \ and\ \bibinfo {author} {\bibfnamefont {L.}~\bibnamefont
  {{Vidmar}}},\ }\bibfield  {title} {\enquote {\bibinfo {title} {{Quantum chaos
  challenges many-body localization}},}\ }\href@noop {} {\bibfield  {journal}
  {\bibinfo  {journal} {arXiv e-prints}\ ,\ \bibinfo {eid} {arXiv:1905.06345}}
  (\bibinfo {year} {2019})},\ \Eprint {http://arxiv.org/abs/1905.06345}
  {arXiv:1905.06345 [cond-mat.str-el]} \BibitemShut {NoStop}%
\bibitem [{\citenamefont {{Halimeh}}\ \emph {et~al.}(2019)\citenamefont
  {{Halimeh}}, \citenamefont {{Yegovtsev}},\ and\ \citenamefont
  {{Gurarie}}}]{Halimeh2019}%
  \BibitemOpen
  \bibfield  {author} {\bibinfo {author} {\bibfnamefont {J.~C.}\ \bibnamefont
  {{Halimeh}}}, \bibinfo {author} {\bibfnamefont {N.}~\bibnamefont
  {{Yegovtsev}}}, \ and\ \bibinfo {author} {\bibfnamefont {V.}~\bibnamefont
  {{Gurarie}}},\ }\bibfield  {title} {\enquote {\bibinfo {title} {{Dynamical
  quantum phase transitions in many-body localized systems}},}\ }\href@noop {}
  {\bibfield  {journal} {\bibinfo  {journal} {arXiv e-prints}\ ,\ \bibinfo
  {eid} {arXiv:1903.03109}} (\bibinfo {year} {2019})},\ \Eprint
  {http://arxiv.org/abs/1903.03109} {arXiv:1903.03109 [cond-mat.stat-mech]}
  \BibitemShut {NoStop}%
\bibitem [{\citenamefont {Efetov}(1983)}]{Efetov1983}%
  \BibitemOpen
  \bibfield  {author} {\bibinfo {author} {\bibfnamefont {K.~B.}\ \bibnamefont
  {Efetov}},\ }\bibfield  {title} {\enquote {\bibinfo {title} {{Supersymmetry
  and theory of disordered metals}},}\ }\href {\doibase
  10.1080/00018738300101531} {\bibfield  {journal} {\bibinfo  {journal}
  {Advances in Physics}\ }\textbf {\bibinfo {volume} {32}},\ \bibinfo {pages}
  {53} (\bibinfo {year} {1983})}\BibitemShut {NoStop}%
\bibitem [{\citenamefont {Argaman}\ \emph {et~al.}(1993)\citenamefont
  {Argaman}, \citenamefont {Imry},\ and\ \citenamefont
  {Smilansky}}]{Argaman1993}%
  \BibitemOpen
  \bibfield  {author} {\bibinfo {author} {\bibfnamefont {N.}~\bibnamefont
  {Argaman}}, \bibinfo {author} {\bibfnamefont {Y.}~\bibnamefont {Imry}}, \
  and\ \bibinfo {author} {\bibfnamefont {U.}~\bibnamefont {Smilansky}},\
  }\bibfield  {title} {\enquote {\bibinfo {title} {{Semiclassical analysis of
  spectral correlations in mesoscopic systems}},}\ }\href {\doibase
  10.1103/PhysRevB.47.4440} {\bibfield  {journal} {\bibinfo  {journal} {Phys.
  Rev. B}\ }\textbf {\bibinfo {volume} {47}},\ \bibinfo {pages} {4440}
  (\bibinfo {year} {1993})}\BibitemShut {NoStop}%
\bibitem [{\citenamefont {Berry}\ and\ \citenamefont
  {Tabor}(1977)}]{Berry1977}%
  \BibitemOpen
  \bibfield  {author} {\bibinfo {author} {\bibfnamefont {M.~V.}\ \bibnamefont
  {Berry}}\ and\ \bibinfo {author} {\bibfnamefont {M.}~\bibnamefont {Tabor}},\
  }\bibfield  {title} {\enquote {\bibinfo {title} {{Level Clustering in the
  Regular Spectrum}},}\ }\href {\doibase 10.1098/rspa.1977.0140} {\bibfield
  {journal} {\bibinfo  {journal} {Proceedings of the Royal Society A:
  Mathematical, Physical and Engineering Sciences}\ }\textbf {\bibinfo {volume}
  {356}},\ \bibinfo {pages} {375} (\bibinfo {year} {1977})}\BibitemShut
  {NoStop}%
\bibitem [{\citenamefont {Fisher}(1995)}]{Fisher1995}%
  \BibitemOpen
  \bibfield  {author} {\bibinfo {author} {\bibfnamefont {D.~S.}\ \bibnamefont
  {Fisher}},\ }\bibfield  {title} {\enquote {\bibinfo {title} {{Critical
  behavior of random transverse-field Ising spin chains}},}\ }\href {\doibase
  10.1103/PhysRevB.51.6411} {\bibfield  {journal} {\bibinfo  {journal} {Phys.
  Rev. B}\ }\textbf {\bibinfo {volume} {51}},\ \bibinfo {pages} {6411}
  (\bibinfo {year} {1995})}\BibitemShut {NoStop}%
\bibitem [{\citenamefont {{Motrunich}}\ \emph {et~al.}(2000)\citenamefont
  {{Motrunich}}, \citenamefont {{Mau}}, \citenamefont {{Huse}},\ and\
  \citenamefont {{Fisher}}}]{Motrunich1999}%
  \BibitemOpen
  \bibfield  {author} {\bibinfo {author} {\bibfnamefont {O.}~\bibnamefont
  {{Motrunich}}}, \bibinfo {author} {\bibfnamefont {S.-C.}\ \bibnamefont
  {{Mau}}}, \bibinfo {author} {\bibfnamefont {D.~A.}\ \bibnamefont {{Huse}}}, \
  and\ \bibinfo {author} {\bibfnamefont {D.~S.}\ \bibnamefont {{Fisher}}},\
  }\bibfield  {title} {\enquote {\bibinfo {title} {{Infinite-randomness quantum
  Ising critical fixed points}},}\ }\href {\doibase 10.1103/PhysRevB.61.1160}
  {\bibfield  {journal} {\bibinfo  {journal} {Phys. Rev. B}\ }\textbf {\bibinfo
  {volume} {61}},\ \bibinfo {pages} {1160} (\bibinfo {year} {2000})},\ \Eprint
  {http://arxiv.org/abs/cond-mat/9906322} {arXiv:cond-mat/9906322
  [cond-mat.dis-nn]} \BibitemShut {NoStop}%
\bibitem [{\citenamefont {{Oshikawa}}(2019)}]{Oshikawa2019UniversalFG}%
  \BibitemOpen
  \bibfield  {author} {\bibinfo {author} {\bibfnamefont {M.}~\bibnamefont
  {{Oshikawa}}},\ }\bibfield  {title} {\enquote {\bibinfo {title} {{Universal
  finite-size gap scaling of the quantum Ising chain}},}\ }\href@noop {}
  {\bibfield  {journal} {\bibinfo  {journal} {arXiv e-prints}\ } (\bibinfo
  {year} {2019})},\ \Eprint {http://arxiv.org/abs/1910.06353} {arXiv:1910.06353
  [cond-mat.stat-mech]} \BibitemShut {NoStop}%
\bibitem [{\citenamefont {Yui}\ and\ \citenamefont {Zagier}(1997)}]{Zagier97}%
  \BibitemOpen
  \bibfield  {author} {\bibinfo {author} {\bibfnamefont {N.}~\bibnamefont
  {Yui}}\ and\ \bibinfo {author} {\bibfnamefont {D.}~\bibnamefont {Zagier}},\
  }\bibfield  {title} {\enquote {\bibinfo {title} {{On the Singular Values of
  Weber Modular Functions}},}\ }\href {\doibase 10.1090/S0025-5718-97-00854-5}
  {\bibfield  {journal} {\bibinfo  {journal} {Math. Comput.}\ }\textbf
  {\bibinfo {volume} {66}},\ \bibinfo {pages} {1645–1662} (\bibinfo {year}
  {1997})}\BibitemShut {NoStop}%
\bibitem [{\citenamefont {{Kudler-Flam}}\ \emph {et~al.}(2019)\citenamefont
  {{Kudler-Flam}}, \citenamefont {{Nie}},\ and\ \citenamefont
  {{Ryu}}}]{Kudler-Flam2019ConformalDiagnostics}%
  \BibitemOpen
  \bibfield  {author} {\bibinfo {author} {\bibfnamefont {J.}~\bibnamefont
  {{Kudler-Flam}}}, \bibinfo {author} {\bibfnamefont {L.}~\bibnamefont
  {{Nie}}}, \ and\ \bibinfo {author} {\bibfnamefont {S.}~\bibnamefont
  {{Ryu}}},\ }\bibfield  {title} {\enquote {\bibinfo {title} {{Conformal field
  theory and the web of quantum chaos diagnostics}},}\ }\href@noop {}
  {\bibfield  {journal} {\bibinfo  {journal} {arXiv e-prints}\ } (\bibinfo
  {year} {2019})},\ \Eprint {http://arxiv.org/abs/1910.14575} {arXiv:1910.14575
  [hep-th]} \BibitemShut {NoStop}%
\bibitem [{\citenamefont {Dyer}\ and\ \citenamefont
  {Gur-Ari}(2017)}]{Dyer2017}%
  \BibitemOpen
  \bibfield  {author} {\bibinfo {author} {\bibfnamefont {E.}~\bibnamefont
  {Dyer}}\ and\ \bibinfo {author} {\bibfnamefont {G.}~\bibnamefont {Gur-Ari}},\
  }\bibfield  {title} {\enquote {\bibinfo {title} {{2D CFT partition functions
  at late times}},}\ }\href {\doibase 10.1007/JHEP08(2017)075} {\bibfield
  {journal} {\bibinfo  {journal} {Journal of High Energy Physics}\ }\textbf
  {\bibinfo {volume} {2017}},\ \bibinfo {pages} {75} (\bibinfo {year}
  {2017})}\BibitemShut {NoStop}%
\bibitem [{\citenamefont {{Di Francesco}}\ \emph {et~al.}(1997)\citenamefont
  {{Di Francesco}}, \citenamefont {Mathieu},\ and\ \citenamefont
  {S{\'{e}}n{\'{e}}chal}}]{DiFrancesco1997}%
  \BibitemOpen
  \bibfield  {author} {\bibinfo {author} {\bibfnamefont {P.}~\bibnamefont {{Di
  Francesco}}}, \bibinfo {author} {\bibfnamefont {P.}~\bibnamefont {Mathieu}},
  \ and\ \bibinfo {author} {\bibfnamefont {D.}~\bibnamefont
  {S{\'{e}}n{\'{e}}chal}},\ }\enquote {\bibinfo {title} {{Modular
  Invariance}},}\ in\ \href {\doibase 10.1007/978-1-4612-2256-9_10} {\emph
  {\bibinfo {booktitle} {Conformal Field Theory}}}\ (\bibinfo  {publisher}
  {Springer New York},\ \bibinfo {address} {New York, NY},\ \bibinfo {year}
  {1997})\ pp.\ \bibinfo {pages} {335--408}\BibitemShut {NoStop}%
\bibitem [{\citenamefont {Benjamin}\ \emph {et~al.}(2019)\citenamefont
  {Benjamin}, \citenamefont {Dyer}, \citenamefont {Fitzpatrick},\ and\
  \citenamefont {Xin}}]{Benjamin2018}%
  \BibitemOpen
  \bibfield  {author} {\bibinfo {author} {\bibfnamefont {N.}~\bibnamefont
  {Benjamin}}, \bibinfo {author} {\bibfnamefont {E.}~\bibnamefont {Dyer}},
  \bibinfo {author} {\bibfnamefont {A.~L.}\ \bibnamefont {Fitzpatrick}}, \ and\
  \bibinfo {author} {\bibfnamefont {Y.}~\bibnamefont {Xin}},\ }\bibfield
  {title} {\enquote {\bibinfo {title} {{The Most Irrational Rational
  Theories}},}\ }\href {\doibase 10.1007/JHEP04(2019)025} {\bibfield  {journal}
  {\bibinfo  {journal} {JHEP}\ }\textbf {\bibinfo {volume} {04}},\ \bibinfo
  {pages} {025} (\bibinfo {year} {2019})},\ \Eprint
  {http://arxiv.org/abs/1812.07579} {arXiv:1812.07579 [hep-th]} \BibitemShut
  {NoStop}%
\bibitem [{\citenamefont {Lau}\ \emph {et~al.}(2019)\citenamefont {Lau},
  \citenamefont {Ma}, \citenamefont {Murugan},\ and\ \citenamefont
  {Tezuka}}]{Lau2018}%
  \BibitemOpen
  \bibfield  {author} {\bibinfo {author} {\bibfnamefont {P.~H.~C.}\
  \bibnamefont {Lau}}, \bibinfo {author} {\bibfnamefont {C.-T.}\ \bibnamefont
  {Ma}}, \bibinfo {author} {\bibfnamefont {J.}~\bibnamefont {Murugan}}, \ and\
  \bibinfo {author} {\bibfnamefont {M.}~\bibnamefont {Tezuka}},\ }\bibfield
  {title} {\enquote {\bibinfo {title} {{Randomness and Chaos in Qubit
  Models}},}\ }\href {\doibase 10.1016/j.physletb.2019.05.052} {\bibfield
  {journal} {\bibinfo  {journal} {Phys. Lett.}\ }\textbf {\bibinfo {volume}
  {B795}},\ \bibinfo {pages} {230} (\bibinfo {year} {2019})},\ \Eprint
  {http://arxiv.org/abs/1812.04770} {arXiv:1812.04770 [hep-th]} \BibitemShut
  {NoStop}%
\bibitem [{\citenamefont {{Young}}\ and\ \citenamefont
  {{Rieger}}(1996)}]{Young1995}%
  \BibitemOpen
  \bibfield  {author} {\bibinfo {author} {\bibfnamefont {A.~P.}\ \bibnamefont
  {{Young}}}\ and\ \bibinfo {author} {\bibfnamefont {H.}~\bibnamefont
  {{Rieger}}},\ }\bibfield  {title} {\enquote {\bibinfo {title} {{Numerical
  study of the random transverse-field Ising spin chain}},}\ }\href {\doibase
  10.1103/PhysRevB.53.8486} {\bibfield  {journal} {\bibinfo  {journal} {Phys.
  Rev. B}\ }\textbf {\bibinfo {volume} {53}},\ \bibinfo {pages} {8486}
  (\bibinfo {year} {1996})},\ \Eprint {http://arxiv.org/abs/cond-mat/9510027}
  {arXiv:cond-mat/9510027} \BibitemShut {NoStop}%
\bibitem [{\citenamefont {Hayn}\ and\ \citenamefont {John}(1987)}]{Hayn1987}%
  \BibitemOpen
  \bibfield  {author} {\bibinfo {author} {\bibfnamefont {R.}~\bibnamefont
  {Hayn}}\ and\ \bibinfo {author} {\bibfnamefont {W.}~\bibnamefont {John}},\
  }\bibfield  {title} {\enquote {\bibinfo {title} {{Effective equations for
  disordered one-dimensional systems}},}\ }\href {\doibase 10.1007/BF01303977}
  {\bibfield  {journal} {\bibinfo  {journal} {Z. Phys. B Condensed Matter}\
  }\textbf {\bibinfo {volume} {67}},\ \bibinfo {pages} {169} (\bibinfo {year}
  {1987})}\BibitemShut {NoStop}%
\bibitem [{\citenamefont {{McKenzie}}(1996)}]{McKenzie1996}%
  \BibitemOpen
  \bibfield  {author} {\bibinfo {author} {\bibfnamefont {R.~H.}\ \bibnamefont
  {{McKenzie}}},\ }\bibfield  {title} {\enquote {\bibinfo {title} {{Exact
  Results for Quantum Phase Transitions in Random XY Spin Chains}},}\ }\href
  {\doibase 10.1103/PhysRevLett.77.4804} {\bibfield  {journal} {\bibinfo
  {journal} {Phys. Rev. Lett.}\ }\textbf {\bibinfo {volume} {77}},\ \bibinfo
  {pages} {4804} (\bibinfo {year} {1996})},\ \Eprint
  {http://arxiv.org/abs/cond-mat/9609195} {arXiv:cond-mat/9609195 [cond-mat]}
  \BibitemShut {NoStop}%
\bibitem [{\citenamefont {{Balents}}\ and\ \citenamefont
  {{Fisher}}(1997)}]{Balents1997}%
  \BibitemOpen
  \bibfield  {author} {\bibinfo {author} {\bibfnamefont {L.}~\bibnamefont
  {{Balents}}}\ and\ \bibinfo {author} {\bibfnamefont {M.~P.~A.}\ \bibnamefont
  {{Fisher}}},\ }\bibfield  {title} {\enquote {\bibinfo {title}
  {{Delocalization transition via supersymmetry in one dimension}},}\ }\href
  {\doibase 10.1103/PhysRevB.56.12970} {\bibfield  {journal} {\bibinfo
  {journal} {Phys. Rev. B}\ }\textbf {\bibinfo {volume} {56}},\ \bibinfo
  {pages} {12970} (\bibinfo {year} {1997})},\ \Eprint
  {http://arxiv.org/abs/cond-mat/9706069} {arXiv:cond-mat/9706069
  [cond-mat.mes-hall]} \BibitemShut {NoStop}%
\bibitem [{\citenamefont {{Song}}\ and\ \citenamefont
  {{Shepelyansky}}(2000)}]{Song1999}%
  \BibitemOpen
  \bibfield  {author} {\bibinfo {author} {\bibfnamefont {P.~H.}\ \bibnamefont
  {{Song}}}\ and\ \bibinfo {author} {\bibfnamefont {D.~L.}\ \bibnamefont
  {{Shepelyansky}}},\ }\bibfield  {title} {\enquote {\bibinfo {title}
  {{Low-energy transition in spectral statistics of two-dimensional interacting
  fermions}},}\ }\href {\doibase 10.1103/PhysRevB.61.15546} {\bibfield
  {journal} {\bibinfo  {journal} {Phys. Rev. B}\ }\textbf {\bibinfo {volume}
  {61}},\ \bibinfo {pages} {15546} (\bibinfo {year} {2000})},\ \Eprint
  {http://arxiv.org/abs/cond-mat/9904229} {arXiv:cond-mat/9904229 [cond-mat]}
  \BibitemShut {NoStop}%
\bibitem [{\citenamefont {Shklovskii}\ \emph {et~al.}(1993)\citenamefont
  {Shklovskii}, \citenamefont {Shapiro}, \citenamefont {Sears}, \citenamefont
  {Lambrianides},\ and\ \citenamefont
  {Shore}}]{Shklovskii1993StatisticsTransition}%
  \BibitemOpen
  \bibfield  {author} {\bibinfo {author} {\bibfnamefont {B.~I.}\ \bibnamefont
  {Shklovskii}}, \bibinfo {author} {\bibfnamefont {B.}~\bibnamefont {Shapiro}},
  \bibinfo {author} {\bibfnamefont {B.~R.}\ \bibnamefont {Sears}}, \bibinfo
  {author} {\bibfnamefont {P.}~\bibnamefont {Lambrianides}}, \ and\ \bibinfo
  {author} {\bibfnamefont {H.~B.}\ \bibnamefont {Shore}},\ }\bibfield  {title}
  {\enquote {\bibinfo {title} {{Statistics of spectra of disordered systems
  near the metal-insulator transition}},}\ }\href {\doibase
  10.1103/PhysRevB.47.11487} {\bibfield  {journal} {\bibinfo  {journal} {Phys.
  Rev. B}\ }\textbf {\bibinfo {volume} {47}},\ \bibinfo {pages} {11487}
  (\bibinfo {year} {1993})}\BibitemShut {NoStop}%
\bibitem [{\citenamefont {{Jacquod}}\ and\ \citenamefont
  {{Shepelyansky}}(1997)}]{Jacquod1997}%
  \BibitemOpen
  \bibfield  {author} {\bibinfo {author} {\bibfnamefont {P.}~\bibnamefont
  {{Jacquod}}}\ and\ \bibinfo {author} {\bibfnamefont {D.~L.}\ \bibnamefont
  {{Shepelyansky}}},\ }\bibfield  {title} {\enquote {\bibinfo {title}
  {{Emergence of Quantum Chaos in Finite Interacting Fermi Systems}},}\ }\href
  {\doibase 10.1103/PhysRevLett.79.1837} {\bibfield  {journal} {\bibinfo
  {journal} {Phys. Rev. Lett.}\ }\textbf {\bibinfo {volume} {79}},\ \bibinfo
  {pages} {1837} (\bibinfo {year} {1997})},\ \Eprint
  {http://arxiv.org/abs/cond-mat/9706040} {arXiv:cond-mat/9706040 [cond-mat]}
  \BibitemShut {NoStop}%
\end{thebibliography}%
\end{document}